\documentclass[12pt]{article}
\usepackage{epsf}
\usepackage{a4}
\usepackage{amsfonts}		% open math symbols
\usepackage{cite}		% collapse citations

\def\hybrid{\topmargin 0pt
        \oddsidemargin 0pt
        \headheight 0pt \headsep 0pt
        \textwidth 6.25in       % A4 paper
        \textheight 9.5in       % A4 paper
        \marginparwidth .875in
        \parskip 5pt plus 1pt   \jot = 1.5ex}

\catcode`\@=11
\def\marginnote#1{}

\newcount\hour
\newcount\minute
\newtoks\amorpm
\hour=\time\divide\hour by60
\minute=\time{\multiply\hour by60 \global\advance\minute by-\hour}
\edef\standardtime{{\ifnum\hour<12 \global\amorpm={am}%
        \else\global\amorpm={pm}\advance\hour by-12 \fi
        \ifnum\hour=0 \hour=12 \fi
        \number\hour:\ifnum\minute<10 0\fi\number\minute\the\amorpm}}
\edef\militarytime{\number\hour:\ifnum\minute<10 0\fi\number\minute}

\def\draftlabel#1{{\@bsphack\if@filesw {\let\thepage\relax
   \xdef\@gtempa{\write\@auxout{\string
      \newlabel{#1}{{\@currentlabel}{\thepage}}}}}\@gtempa
   \if@nobreak \ifvmode\nobreak\fi\fi\fi\@esphack}
        \gdef\@eqnlabel{#1}}
\def\@eqnlabel{}
\def\@vacuum{}
\def\draftmarginnote#1{\marginpar{\raggedright\scriptsize\tt#1}}

\def\draft{\oddsidemargin -.5truein
        \def\@oddfoot{\sl preliminary draft \hfil
        \rm\thepage\hfil\sl\today\quad\militarytime}
        \let\@evenfoot\@oddfoot \overfullrule 3pt
        \let\label=\draftlabel
        \let\marginnote=\draftmarginnote
   \def\@eqnnum{(\theequation)\rlap{\kern\marginparsep\tt\@eqnlabel}%
\global\let\@eqnlabel\@vacuum}  }

\def\numberbysection{\@addtoreset{equation}{section}
        \def\theequation{\thesection.\arabic{equation}}}

\def\titlepage{\@restonecolfalse\if@twocolumn\@restonecoltrue\onecolumn
     \else \newpage \fi \thispagestyle{empty}\c@page\z@
        \def\thefootnote{\fnsymbol{footnote}}
	\setcounter{page}{0} }
\def\endtitlepage{\if@restonecol\twocolumn \else  \fi
        \def\thefootnote{\arabic{footnote}}
        \setcounter{footnote}{0}}  %\c@footnote\z@ }

\catcode`@=12
\relax

\def\ie{\hbox{\it i.e.}}

\def\beq{\begin{equation}}
\def\eeq{\end{equation}}
\def\bea{\begin{eqnarray}}
\def\eea{\end{eqnarray}}
\relax
\numberbysection
\hybrid
\begin{document}
\begin{titlepage}
\begin{center}
{\large\bf
% On the 2-D random-bond Ising model.
Strong disorder fixed points in the two-dimensional random-bond Ising model
}\\[.3in] 

{\bf M.\ Picco$^{1}$, A.\ Honecker$^{2}$ and P.\ Pujol$^{3}$}\\
        % {\bf (1)}
	$^1$ {\it LPTHE\/}\footnote[1]{Unit\'e mixte de recherche du CNRS 
UMR 7589.}, % \\
        {\it  Universit\'e Pierre et Marie Curie-Paris6 and \\
              Universit\'e Denis Diderot-Paris7\\
              Bo\^{\i}te 126, Tour 24-25, 5 \`eme \'etage, \\
              4 place Jussieu,
              F-75252 Paris CEDEX 05, France, \\
    e-mail: {\tt picco@lpthe.jussieu.fr}. }\\
        % {\bf (2)}
	$^2$ {\it Institut f\"ur Theoretische Physik, TU Braunschweig,\\
    Mendelssohnstr.\ 3, 38106 Braunschweig, Germany, and \\
    Institut f\"ur Theoretische Physik, Universit\"at G\"ottingen, \\
    Friedrich-Hund-Platz 1, 37077 G\"ottingen, Germany, \\
    e-mail: {\tt a.honecker@tu-bs.de}. } \\
        %{\bf (3)}
	$^3$ {\it Laboratoire de Physique\footnote[3]{Unit\'e mixte de 
             recherche du CNRS UMR 5672 
            associ\'ee \`a l'Ecole Normale Sup\'erieure de Lyon.}, % \\
             ENS Lyon, \\
             46 All\'ee d'Italie, 69364 Lyon C\'edex 07, France, \\
    e-mail: {\tt Pierre.pujol@ens-lyon.fr}. }\\
\end{center}
%\vskip .04in
\centerline{(Dated: June 10, 2006)}
\vskip .2in
\centerline{\bf ABSTRACT}
\begin{quotation}
%{\bf Abstract.}
The random-bond Ising model on the square lattice has several
disordered critical points, depending on the probability distribution
of the bonds. There are a finite-temperature multicritical point,
called Nishimori point, and a zero-temperature fixed point, for both a
binary distribution where the coupling constants take the values $\pm
J$ and a Gaussian disorder distribution. Inclusion of dilution in the
$\pm J$ distribution ($J=0$ for some bonds) gives rise to another
zero-temperature fixed point which can be identified with percolation
in the non-frustrated case ($J \ge 0$).  We study these fixed points
using numerical (transfer matrix) methods. We determine the location,
critical exponents, and central charge of the different fixed points
and study the spin-spin correlation functions. Our main findings are
the following:
(1) We confirm that the Nishimori point is universal with respect to the type of
disorder, \ie\ we obtain the same central charge and critical exponents
for the $\pm J$ and Gaussian distributions of disorder.
(2) The Nishimori point, the zero-temperature fixed point for the $\pm
J$ and Gaussian distributions of disorder, and the percolation point
in the diluted case all belong to mutually distinct universality
classes.
(3) The paramagnetic phase is re-entrant below the Nishimori point, \ie\
the zero-temperature fixed points are not located exactly below the
Nishimori point, neither for the $\pm J$ distribution, nor for the
Gaussian distribution.
\vskip 0.5cm 
\noindent
%\pacs
%PACS numbers: 05.70.Jk, 64.60.Ak, 64.60.Fr

% PACS 1999 classification: http://www.aip.org/pacs/pacs99/pacscheme.html

\end{quotation}
\end{titlepage}
\section{Introduction}

The problem of disordered magnetic systems has attracted great
interest in the past years and many questions still remain
unanswered. An interesting problem is the universality class of second
order phase transitions in two-dimensional systems. The random-bond
Ising model (RBIM) is one of the simplest and best known of these
systems \cite{DDSL,N,ON,LG}, but exhibits rich enough behaviour to
give a general understanding of the problem. Further interest in the
RBIM stems from analogies with the quantum Hall transition
\cite{CF,GRL,CRKHAL,MENMD06} and applications in coding theory
\cite{Sourlas,NishCod,Iba,Preskill}.

For a small amount of randomness, the universality class of the RBIM
remains unchanged, presenting only logarithmic corrections in some
correlation functions \cite{DDSL}. For some particular kind of
randomness, namely dilution, one can show that another non-trivial
fixed point corresponds to a percolation universality class at zero
temperature, since it becomes a purely geometric problem of having a
thermodynamic number of spins within the same cluster.  It is
important to notice that in this case, all the randomly distributed
bonds are non-negative.

The situation is quite different if some negative bonds are allowed in
the probability distribution. For certain distributions with negative
bonds, Nishimori \cite{N, ON} has shown that some exact statements can
be made about physical quantities. There is in particular a line in
which the internal energy can be calculated exactly, known as the
Nishimori line. The interest of this line goes further, since it has
been shown that this line is invariant under renormalization group
(RG) transformations \cite{LG}. Since this lines crosses the
ferromagnetic to paramagnetic transition line, the intersection point,
known as the Nishimori point is a fixed point. Examples of probability
distributions satisfying the Nishimori condition are the Gaussian,
$\pm J$ binary and $0, \pm J$ with appropriately chosen weights (see
below). Despite recent analytical approaches \cite{GRL}, the exact
characterization of the universality class of this non-perturbative
fixed point is still unknown. The underlying field theory describing
this point is certainly a good representative of the disordered
fixed-point behaviour with very interesting and rich phenomenology
\cite{Z,GRLC}. In a previous letter \cite{HPP}, we have studied
numerically the critical exponents and central charge of the Nishimori
point in the $\pm J$ RBIM.  We found in particular that its
universality class does not correspond to the one of percolation, as
one could have imagined in view of earlier numerical investigations of
this model \cite{MM,SA,AQdS}.  This is derived, first, from numerical
estimates for the critical exponents and central charge which differ
significantly from those of percolation, and confirmed by an analysis
of higher moments of the correlation functions (which are all equal in
percolation and only in pairs for the Nishimori point, see \cite{HPP}
for details and later in this paper).

The conclusion of a different universality class was confirmed in
another numerical analysis by Merz and Chalker \cite{MC}. Thanks to a
mapping to a network model, these authors reached big lattice sizes
and high accuracy in the measurement of critical exponents. An
interesting remark made by these authors concerns the dual theory of
the RBIM\footnote{Because of the randomness introduced in the bonds,
the model is not self-dual as the pure Ising model.}, in which the
different moments of the disorder field acquire negative dimensions
\cite{MC2}. More recently, Nishimori and Nemoto \cite{NN} used a
generalized model with self-duality, and conjectured that the
projection onto the RBIM gives the phase boundary for this model. This
result permits in particular to locate the Nishimori point analytically.
The conjectured location is however outside the accuracy range
of the most recent numerical works on the $\pm J$ binary disorder case
\cite{HPP,MC} and the validity of this conjecture is certainly a very
interesting open issue.

%\begin{table}[hpt]
\begin{table}
\centerline{\begin{tabular}{|c|c|c|}
\hline
$p_c$               & Method            & Reference \\ \hline
\multicolumn{3}{|c|}{Nishimori point in the $\pm J$ model} \\ \hline
$0.111 \pm 0.002$   & Transfer matrix   & \cite{OzekiNishimori} \\
$0.114 \pm 0.003$   & Series expansion  & \cite{SA} \\
$0.1128 \pm 0.0008$ & Non-equilibrium   & \cite{OzekiIto} \\
$0.1095 \pm 0.0005$ & Transfer matrix   & \cite{AQdS} \\
$0.1094 \pm 0.0002$ & Transfer matrix   & % This work, see also
 \cite{HPP} \\
$0.1093 \pm 0.0002$ & Fermionic transfer matrix & \cite{MC} \\
$0.110028$          & Duality           & \cite{NN} \\
$\le 0.178203$      & Rigorous upper bound & \cite{MNN03} \\ \hline 
\multicolumn{3}{|c|}{$T=0$ critical point in the $\pm J$ model} \\ \hline
$ \sim 0.099$ & Series expansion & \cite{Grinstein} \\
$ 0.105 \pm 0.01$ & Matching algorithm & \cite{Freund} \\
$ 0.095<p_c < 0.108$ &
% Efficient
Matching algorithm & \cite{Bendish} \\
${\textstyle 0.104 \pm 0.001 \atop \textstyle
0.106 \pm 0.002}$ & Exact ground states & \cite{KaRi} \\
$0.115$             & Ground state enumeration & \cite{BGP} \\
$0.1031 \pm 0.0001$   & Exact ground states & \cite{Preskill} \\
$0.103 \pm 0.001$   & Exact ground states & \cite{Hartmann} \\ \hline
\end{tabular}
}
\caption{
Overview of estimates for $p_c$ at fixed points in the two-dimensional
$\pm J$ random-bond
Ising model. The first part of this table is for the Nishimori point,
the second part for the zero-temperature critical point.
\label{sumPc}
}
\end{table}

\begin{figure}
\centerline{\epsfxsize=300pt{\epsffile{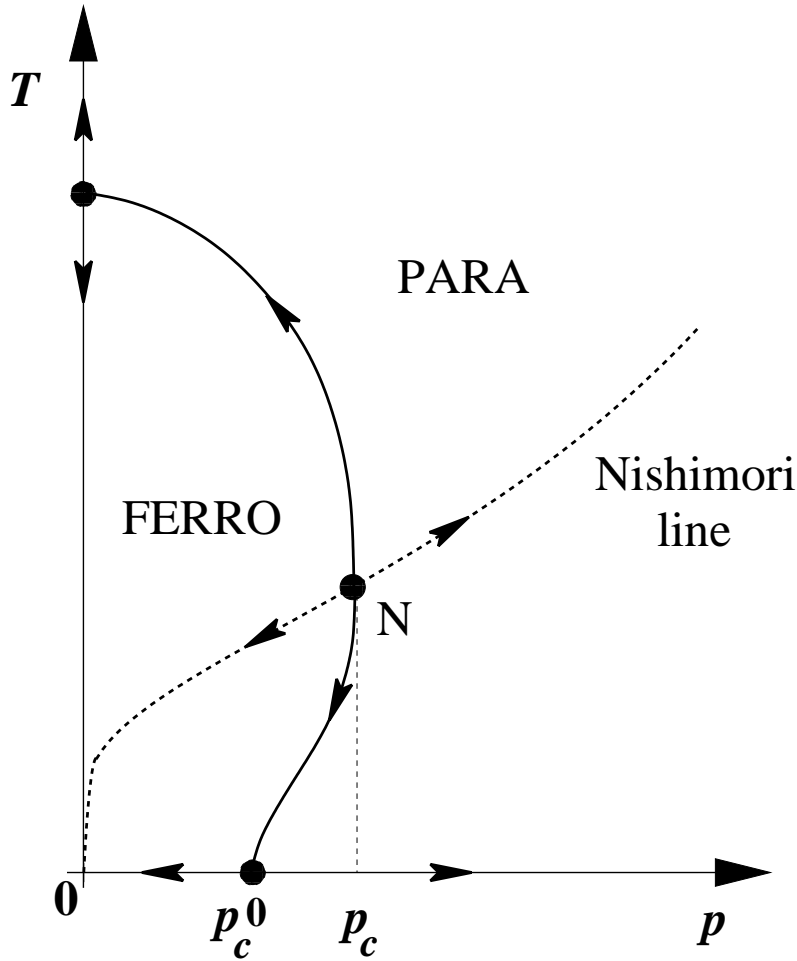}}}
\protect\caption[2]{\label{pd} Phase diagram of the two-dimensional
$\pm J$ random-bond Ising model.
The arrows represent the flow under the renormalization group.} 
\end{figure}

The pure and the Nishimori point, when present in the phase diagram,
are not the only non-trivial fixed points of the model: there is in
any case a zero-temperature fixed point. While Nishimori's results
state rigorously that this point cannot be located at a higher
concentration of ``impure'' bonds, analytical and numerical works on
the $\pm J$ model
\cite{Grinstein,Freund,Bendish,Preskill,Hartmann,KaRi,MB,BGP} tend to
conclude that it is located at a smaller density of impurities,
indicating a re-entrance of the ferromagnetic phase. The properties of
the zero-temperature point vary considerably with the kind of disorder
introduced. For a symmetric Gaussian distribution, it has been shown
\cite{NS} that the lowest energy configuration is unique (modulo the
${\mathbb Z}_2$ symmetry) with probability one. One is tempted to
check the extension of this result to our zero-temperature point, and
check for example that, for a given configuration of the disorder, any
spin-spin correlation function is $1$ or $-1$. Then, all the odd
moments of the spin-spin correlation functions are equal, and all the
even moments are just equal to $1$, a result that is similar, but not
exactly identical to the percolation case.  The situation is much more
subtle for distributions like the $\pm J$ one, since frustration plays
a crucial role. For a generic configuration of disorder, the lowest
energy states are expected to be highly degenerate. The results
obtained by different techniques for the location of both, the
Nishimori point and the zero-temperature critical point are summarized
in Table~\ref{sumPc} for the $\pm J$ model. As can be seen from the
most recent results, the zero-temperature critical point seems to be
located at a concentration of ``impure'' bonds strictly smaller than
the one of the Nishimori point.  A schematic picture of the phase
diagram of the $\pm J$ case in the $p-T$ plane ($p$ being the number
of antiferromagnetic bonds) is shown in Fig.~\ref{pd}, where the
separation of the paramagnetic and ferromagnetic phases, as well as
the Nishimori line are drawn.  The location of the Nishimori point N
and the zero-temperature critical point are representative of the
results shown in Table~\ref{sumPc}. For the case of the Gaussian
disorder, a similar diagram may be obtained by replacing the parameter
$p$ with the variance of the distribution of disorder, $\sigma$,
although the shape of the Nishimori line is different.

%Moreover, this point seems to share
%the same critical exponents obtained for the finite temperature
%Nishimori point.

Given a simple phase diagram as shown in Fig.~\ref{pd}, it is tempting
to trace the massless flow backwards from the pure Ising model at $p=0$
to the Nishimori point N \cite{MS95,CHMP}. However, we will argue in
this paper that in a generalized parameter space there are in fact several
fixed points on the massless surface with different universality classes
such that a backtracking procedure is in general ambiguous. Indeed,
the problems encountered in \cite{CHMP} were one of the original
motivations for our earlier numerical study of the Nishimori point \cite{HPP}
which we continue and expand here.

In this paper we provide an extensive numerical analysis of the
random-bond Ising model for three different kinds of probability distributions. 
We first consider in parallel the case of Gaussian and $\pm J$ distributions
for which we estimate the location of the
fixed point along the Nishimori line using domain-wall techniques.
We compute the free energy with a transfer matrix technique \cite{Night}
and obtain the central charge by analyzing the finite-size corrections.
We also compute the spin-spin
correlation functions on long strips to obtain the magnetic exponent. This and other
measurement to check consistency clearly show that the models with $\pm J$ and Gaussian 
distributions share the same universality class at their respective Nishimori points.
The compatibility of the results between the $\pm J$ and Gaussian cases
further elucidates the nature of the universality class of this point which
was recently argued to be different from the one of percolation \cite{HPP,MC}.
We next consider dilution, which can be modeled by a probability
distribution allowing the values of $\pm J$ and $0$ for the coupling
constants. In the
case of pure dilution (allowing only values $+J$ and $0$ for the
coupling constants), the other fixed point apart the one of the pure model is located 
at zero temperature and dilution $q_c = 1/2$ and simply corresponds to
bond percolation. 
The $\pm J$, $0$ distribution allows us to study the behaviour in the critical
line between the Nishimori point and the percolation point. 
In particular, by studying the value of the effective central charge for different strip widths 
and its evolution towards larger sizes we show that the percolation fixed 
point is repulsive along this critical line in favour of the
Nishimori point. 
The same technique is used to confirm that the Nishimori point is unstable with respect
to the pure Ising fixed point when moving on the critical line connecting these two points.
We finally address the problem of the zero-temperature fixed point.
Although for the case of the $\pm J$ distribution there is an extensive list of numerical
works indicating re-entrance of the ferromagnetic phase
(see Table~\ref{sumPc}), to our knowledge there was
no conclusive evidence of the same fact for the Gaussian distribution.
We show here clear 
evidence for the re-entrance of the phase also for the Gaussian distribution.  
We compute also
different moments of spin-spin correlation functions and the magnetic exponent. Our results
in the magnetic sector clearly show that the universality class of this
zero-temperature critical point is once again
different from the one of the finite-temperature Nishimori point as well
as percolation, in contrast to what one could have thought considering previous numerical
results in two dimensions \cite{SA,KaRi}.   

\section{Some definitions}

In this section, we present some definitions which will be used
throughout this paper. We will consider two kinds of probability
distributions $P(J)$ for the bonds: a discrete distribution where the
coupling constants can take values $\pm 1$ and $0$, and a continuous Gaussian
distribution. In the case of a discrete distribution, two subclasses
can be considered, the first one allows only the
values $\pm 1$ while the second and more general one, allows also
the value $0$ for the coupling constants. This last case corresponds to
dilution. In any case we can imagine a phase diagram in which the
vertical axis is given by temperature and the horizontal axis by a parameter
representing the strength of the disorder (see Figure \ref{pd}). 

Let us define the Hamiltonian of the two-dimensional Ising model to be:
\beq
\label{HIsing}
H = - \sum_{\langle i,j \rangle} J_{i,j} \, \delta_{S(i),S(j)} \, ,
\eeq
were $S(i) = \pm 1$ and $\langle i,j \rangle$ means nearest neighbours on
a square lattice. The variables $J_{i,j}$ are random and independently
chosen with a probability distribution function $P(J)$. As usual, the Kronecker 
delta function is zero if the spins are different and one if they are equal. 
To establish the relation to other conventions for
the energy note that it can be expressed in terms of products of Ising
spins via $\delta_{S(i),S(j)} = \left(S(i)\,S(j)+1\right)/2$.

The Nishimori line is defined % in this phase diagram
by the condition \cite{N}:
\beq
P(-J)= \exp(- \beta J) \, P(J) \, ,
\label{nishcond}
\eeq
where $\beta = {1 \over k_{B} T}$ is the inverse temperature
(from now on we choose the convention $k_{B} =1$). 
Let us analyze the Nishimori condition for the different distributions
considered here
\begin{itemize}
\item $J=\pm 1$: The distribution is characterized by the concentration of
antiferromagnetic bonds $p$:
\beq
P(J)=(1-p)\,\delta(J-1)+p\,\delta(J+1) \; .
\eeq
Eq.~(\ref{nishcond}) gives the following condition for the Nishimori line
\beq
\beta=\ln{\left( 1-p \over p \right)} \; ,
\label{pmJnis}
\eeq
which is schematically depicted in Fig.~\ref{pd}. The line extends
from the (attractive) fixed
points given by $T=0$, $p=0$ and $T = \infty$, $p=1/2$ and crosses
the Para-Ferro transition
line at the critical concentration $p_c$. In the first part
of Table~\ref{sumPc} we summarize estimates for the location
of the critical point $p_c$.  With the exception of a conjectured duality
property \cite{NN} and a rigorous upper bound \cite{MNN03},
all the other results come from numerical simulations. 
\item Gaussian distribution:
\beq
\label{gaussdis}
P(J) = \sqrt{ 1 \over 2 \pi \sigma^2} \,
\exp{\left(-{(J-J_0)^2\over 2 \sigma^2}\right)} \; ,
\eeq
with the following condition
\beq
\beta= {J_0\over \sigma^2}
\eeq
for the Nishimori line. In the following, we will choose $J_0=1$
without any loss of generality. Thus, the distribution is
characterized by the parameter $\sigma$ along the Nishimori line which
extends from the (attractive) fixed points given by $T=0$, $\sigma=0$
and $T = \infty$, $\sigma=\infty$ and crosses the Para-Ferro transition
line at the value $\sigma_c$. The first numerical characterization of
the critical point was given in \cite{MM}, with a value of $\sigma_c
\sim 0.97$.
\begin{figure}
\centerline{\epsfxsize=300pt{\epsffile{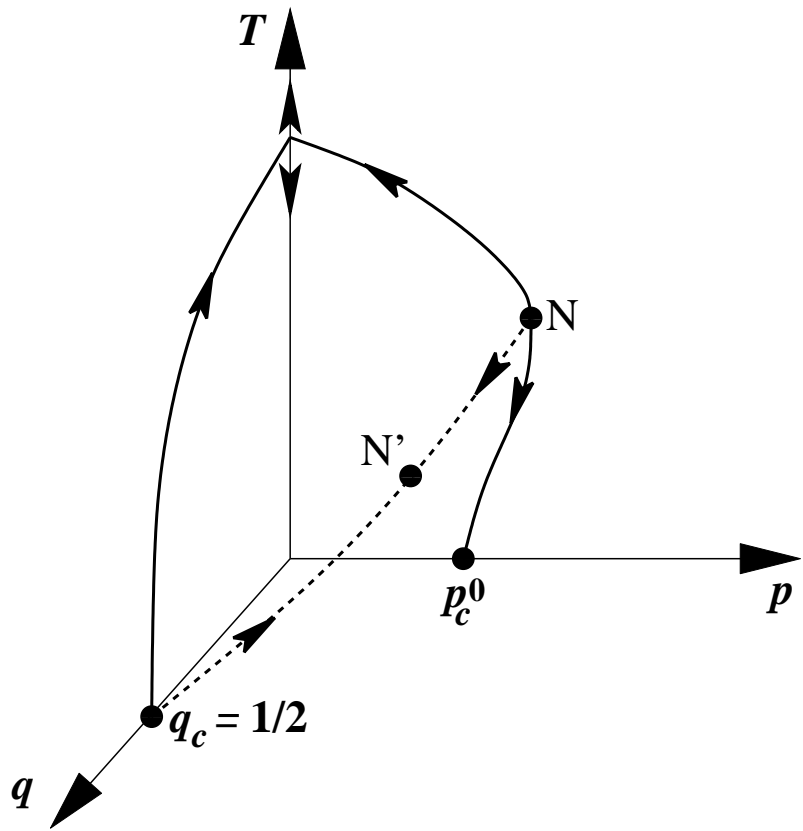}}}
\protect\caption[2]{\label{pdbin} Phase diagram of the diluted $\pm
J,~0$ random-bond Ising model. $p$ and $q$ are the concentration of
negative and zero bonds, respectively. The arrows represent the flow
under the renormalization group.
The dashed line represents the intersection between the Nishimori
surface and the Ferro-Para transition surface.} 
\end{figure}
\item Finally, the binary distribution can be generalized to include
dilution. In this case, we write:
\beq
P(J)=q \, \delta(J) + (1-q-p) \, \delta (J-1) + p \, \delta(J+1) \; .
\eeq
The case $q=0$ corresponds to the binary distribution discussed above,
while $p=0$ gives the ferromagnetic diluted model. In the diluted
model there is only a fixed point located at zero temperature and $q_c
= 1/2$, apart from the pure model fixed point \cite{YeSt79}.
This $T=0$ fixed point corresponds to percolation.
The reason for this is that in the absence of frustration
all the spins belonging to the same cluster must point to the same direction
at zero temperature. Then, whenever there is percolation, or a cluster containing a 
macroscopic number of spins, there is a macroscopic magnetization,
and this happens until the critical concentration of bonds $1/2$
(see for example \cite{YeSt79,StAh}).

The Nishimori surface in the $T$-$p$-$q$ space is now given by:
\beq
\label{nissurface}
\beta= \ln{\left( 1-p-q \over p \right)} \; .
\eeq
The intersection of this surface with the ferromagnetic-paramagnetic
transition surface gives a renormalization group invariant line
\cite{LG} whose end points are the percolation $q=1/2,$ $T=0$ fixed
point and the finite-temperature Nishimori point N of the purely
binary case as depicted in Fig.~\ref{pd}.
In the $T-p-q$ phase diagram, the location of the Nishimori point,
here called N', must be within the intersection line mentioned above
(the dashed line in Fig.~\ref{pdbin}).
In this sense, the point N depicted in Fig.~\ref{pd} is the representative
in the $T-p$ space of the more general location of the Nishimori point
denoted by N' in Fig.~\ref{pdbin}. 
\end{itemize}

\section{Domain-wall free energy}

\label{secDW}

In the following section we will discuss the domain-wall free energy
in a manner very similar to \cite{MM}.
For a strip of with $L$ the domain-wall free energy $d_L$ is defined as
\beq
d_L = L^2 \left(f_L^{(p)} - f_L^{(a)}\right) \, ,
\label{defDW}
\eeq
where $f_L^{(p)}$ is the free energy {\it per site} $f_L^{(p)} = {\ln
Z^{(p)} \over L M}$ of a strip of width $L$ and length $M$ with {\it
periodic} boundary conditions\footnote{Note that our sign conventions
differ from the standard ones.} and $f_L^{(a)} = {\ln Z^{(a)} \over L
M}$ the corresponding one with {\it antiperiodic} boundary conditions.

$d_L$ measures the free energy associated to a domain wall in the
system. We will first consider the $\pm J$ distribution of
disorder. For fixed parameters $p$ ($\beta$) in the disordered
(paramagnetic) phase, one should have $\lim_{L \to \infty} d_L \to 0$
while $d_L$ should diverge in the ordered phase ($d_L \to \infty$ as
$L \to \infty$). At the fixed point $p_c$ ($\beta_c$), $d_L$ should
converge quickly with $L$.  We can therefore use crossing points
between $d_{L_1}$ and $d_{L_2}$ to obtain finite-size estimates for
$p_c$.

Let us consider first the $\pm J$ distribution of disorder.
$f_L^{(p)}$ and $f_L^{(a)}$ are computed with the transfer matrix
technique (see, e.g., \cite{Night})
on strips of length $10^6$. Averages over up to $N \approx
4000$ samples of such $L \times 10^6$ strips are taken in order to
average over randomness and to determine the statistical error.  It is
useful to fix the number of bonds on each $L \times 10^6$ strip to
approximate the chosen value of $p$ as accurately as possible: In our
implementation such a strip has $L \cdot (2 \cdot 10^6 -1)$ bonds out
of which we select the integer closest to $p \, L \cdot (2 \cdot 10^6
-1)$. We found that this method leads to 5 times smaller error bars for
$f_L^{(p)}$ than if one selects each bond {\it separately} at random with
probability $p$, {\it i.e.}\ for the same precision one needs 25 times
less samples with this method compared to generating each bond at
random with probability $p$ without constraining their total number.
For $d_L$ the improvement is even bigger and the error bars become 10 times
smaller -- or one needs 100 times less samples for the the same accuracy. 
\begin{figure}[ht]
\begin{center}
\epsfxsize=400pt{\epsffile{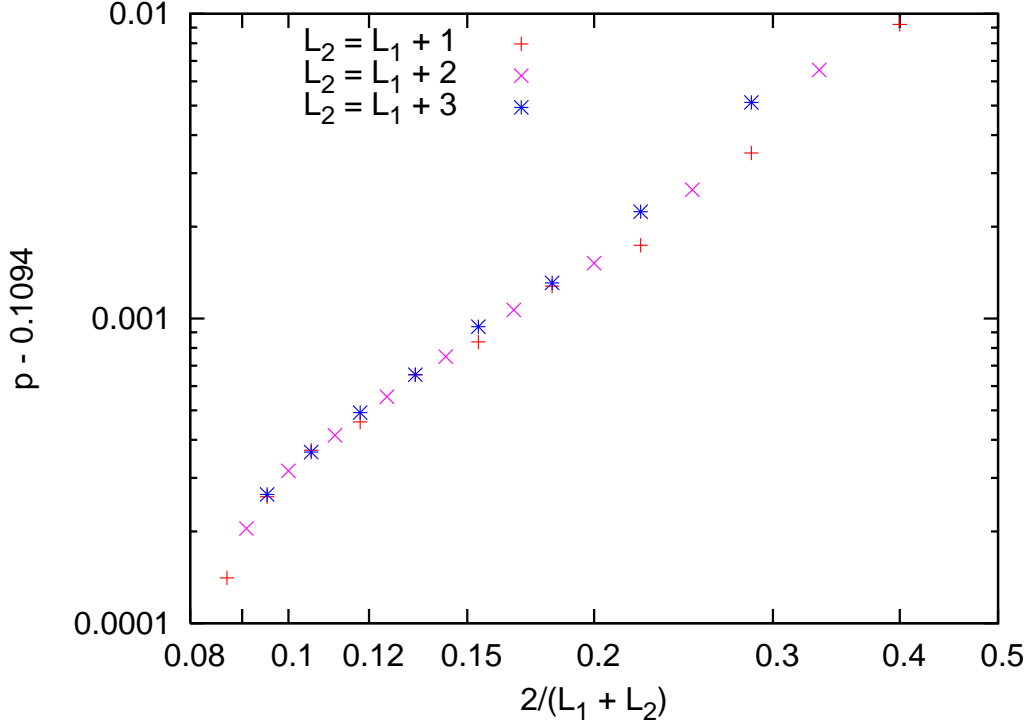}}
\end{center}
\caption{
Position $p$ of crossings between $d_{L_1}(p)$ and $d_{L_2}(p)$
on the Nishimori line of
the $\pm J$ random-bond Ising model with $L_2 \le 12$.
\label{figCross}
}
\end{figure}

Next we will present some details of the analysis leading to the results
presented in \cite{HPP} -- a very similar computation was also performed
in \cite{MC}. Fig.\ \ref{figCross} shows a doubly logarithmic plot
of the crossing point $p_c(L_1,L_2)$ between $d_{L_1}(p)$ and
$d_{L_2}(p)$, % determined from the new data.  anticipating already
using $p_c = 0.1094$ in the shift of the vertical axis.  It is known for the
pure Ising model that the finite-size corrections to $p_c$ scale
roughly $\sim L^{-2}$ with $L=(L_1+L_2)/2$ \cite{Sor}. In order to
extrapolate $p_c$ to the thermodynamic limit, one can therefore first
use a plot of finite-size estimates for $p_c$ as a function of
$L^{-2}$ and extrapolate to the vertical axis. A reasonable
extrapolation with a generous error bar is
\beq
p_c = 0.1093 \pm 0.0004 \, .
\label{pcrit}
\eeq           
Even if the correction is probably not of the form $1/L^2$ in the present
case, one can see in Fig.\ \ref{figCross} that the finite-size correction
to $p_c$ is well described by a power in $L$. For an improved extrapolation
we therefore use the following form for a fit:
\beq
p_c(L_1,L_2) = p_c + \alpha \left({L_1 + L_2 \over 2}\right)^{-\xi} \, .
\label{pcritFit}
\eeq
We then find \cite{HPP}
\beq
p_c = 0.1094 \pm 0.0002
\label{pcrit1}
\eeq
and an exponent
\beq
\xi = 1.5 \pm 0.3 \, .
\label{xiEst}
\eeq
The two estimates (\ref{pcrit}) and (\ref{pcrit1}) agree well with
each other -- the error bar of the second one is just a little smaller.
The exponent (\ref{xiEst}) cannot be determined very accurately.

\begin{figure}[ht]
\begin{center}
\epsfxsize=400pt{\epsffile{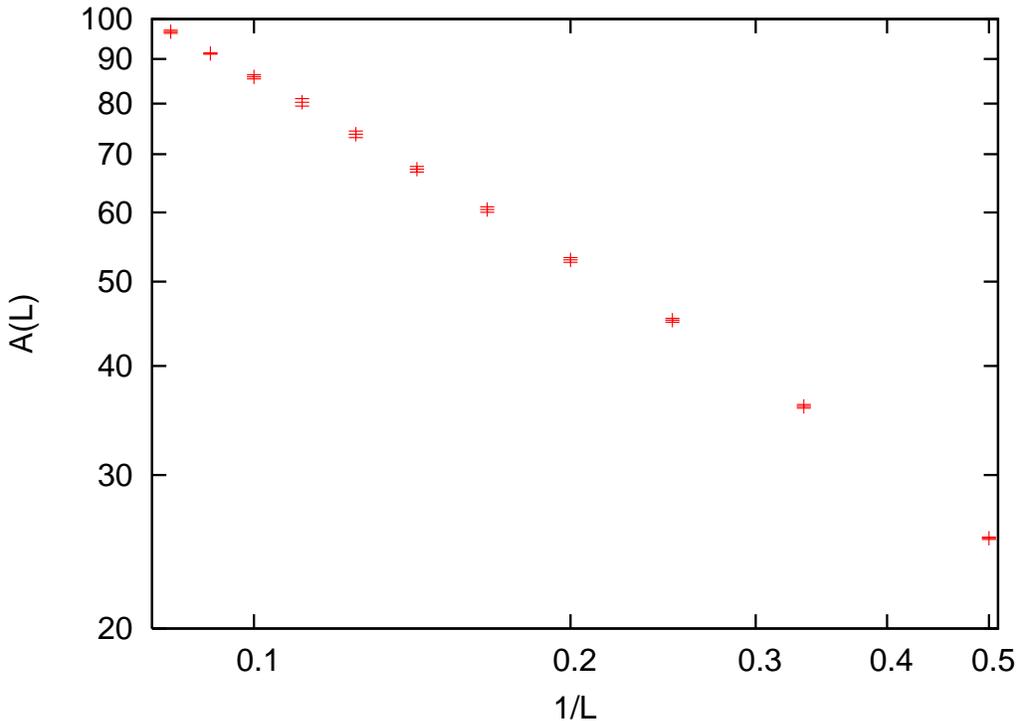}}
\end{center}
\caption{
Slope $A(L)$ of the domain-wall free energy
for the $\pm J$ random-bond Ising model as a function of $1/L$.
\label{figSlopes}
}
\end{figure}

The estimates (\ref{pcrit}) and (\ref{pcrit1}) should be compared with
other estimates which are summarized in Table~\ref{sumPc}. At this
point comparison with the zero-temperature transition is meaningful
only if one assumes that the $T=0$ transition point is located
exactly below the Nishimori point. Our estimate agrees within error
bars with other transfer matrix computations
\cite{OzekiNishimori,AQdS,MC}.
%-- (\ref{pcrit1}) is actually very close to that of \cite{AQdS},
%just with a reduced error bar.
However, our estimate falls outside error bars of estimates obtained
by other methods: It is smaller than the estimates of
\cite{SA,OzekiIto} (in the latter case significantly).
% and only
%the estimates of \cite{KaRi} are smaller than ours (though it remains
%to be clarified to which extent the latter are relevant to the
%Nishimori point).
This discrepancy % with \cite{SA,OzekiIto}
becomes clearest if one notices that several of our finite-size
crossings have already reached the region $p \approx 0.1097$
(see also Fig.\ 2 of \cite{HPP}). If one
now assumes that these crossings decrease monotonically with
increasing $L$ (as the data in Fig.\ \ref{figCross} indeed does), one
does not need to actually carry out the extrapolation and estimate its
error in order to conclude that our extrapolated value for $p_c$ must
end up below the error windows of \cite{SA,OzekiIto}.
At this point it is particularly reassuring to observe that
our result is fully consistent with the one of \cite{MC} which
has been obtained using also substantially wider strips (up to $L=64$).
Furthermore, (\ref{pcrit1}) is in complete agreement with 
other quantities to be discussed in later sections. Thus, we
observe a significant difference between the analytic estimate
of \cite{NN} obtained from a duality argument and our value for $p_c$
as well as the one of Merz and Chalker \cite{MC}.

Now let us look at a first critical exponent, namely the
correlation length exponent $\nu$ along the Nishimori line. If one assumes
the scaling form
\beq
d_L(p-p_c) = d\left((p-p_c) L^{1/\nu} \right) \, ,
\label{dScal}
\eeq
one can expand around $p_c$ and finds
\beq
d_L(p - p_c) \sim \hbox{const.} - A(L)\, p
\label{dAsym}
\eeq
with
\beq
A(L) \sim L^{1/\nu} \, .
\label{nuFit}
\eeq
Then one can fit $d_L$ close to $p_c$ by a linear function and extract $A(L)$.

Fig.\ \ref{figSlopes} shows the values for $A(L)$ determined in this manner
on a doubly logarithmic scale\footnote{Our data for $d_L$ is in
perfect agreement with that of \cite{MC} where we overlap. However,
the windows in $p$ used for the present estimates of $A(L)$ differ from
those used in \cite{HPP} leading to slightly different results.
}. One can see that they follow indeed a power law.
Using (\ref{nuFit}) we extract
\beq
\nu = 1.48 \pm 0.03 \, ,
\label{nuVal}
\eeq
which amounts to a slight correction of the value $\nu = 1.33 \pm 0.03$
obtained in \cite{HPP}. The result (\ref{nuVal}) is now in excellent
agreement with the value $\nu = 1.50 \pm 0.03$
obtained by a fermionic transfer matrix \cite{MC}.
%Note that using a transfer matrix method McMillan \cite{MM} obtained
%$\nu = \lambda_1^{-1} \approx 1.58$ for the Gaussian random-bond Ising model.
The most recent results $\nu \approx 1.5$ do no longer agree
well with $\nu = 1.32 \pm 0.08$ obtained by high-temperature series \cite{SA}.
However, we already observed above that the series expansion method
does not yield a very accurate estimate for $p_c$ either. Furthermore,
the value $\nu = 4/3$ characteristic for percolation (see e.g.\ \cite{StAh})
now falls outside numerical errors and thus it
is possible to conclude that the Nishimori point is not in the universality
class of percolation already on the basis of the exponent $\nu$.
Note that the result (\ref{nuVal}) does not involve locating $p_c$ precisely
and should therefore be independent of errors which may have been made in the
location of $p_c$.

\begin{figure}
\begin{center}
\epsfxsize=400pt{\epsffile{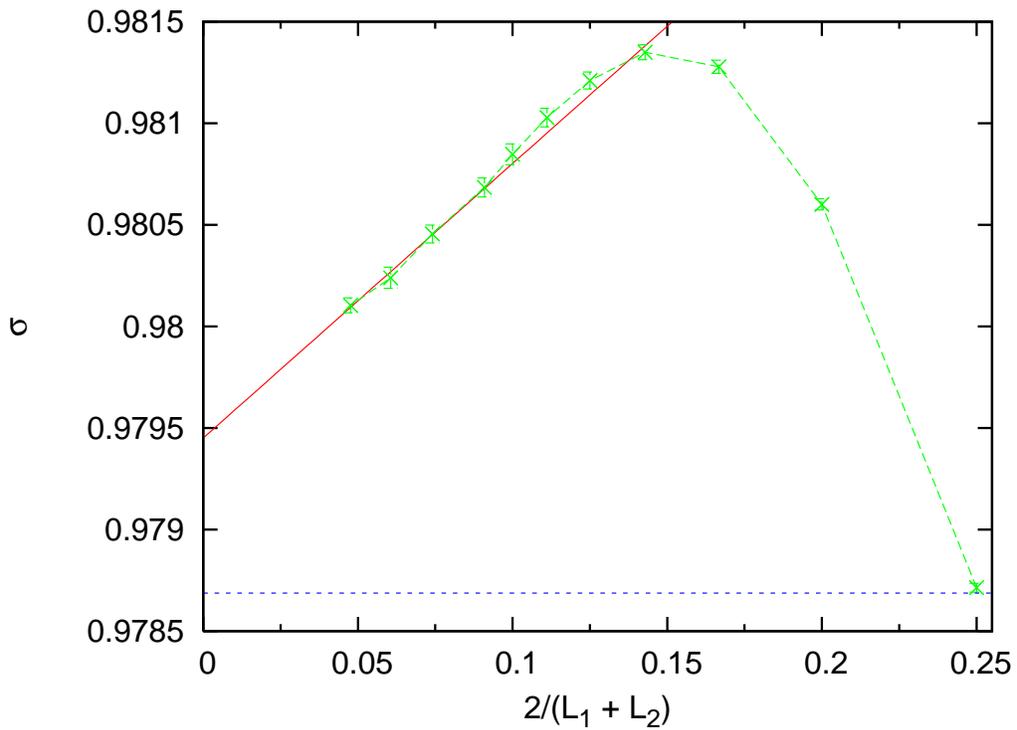}}
\end{center}
\protect\caption[2]{\label{crossg}Position $\sigma$ of crossings
between $d_{L_1}(\sigma)$ and $d_{L_2}(\sigma)$
on the Nishimori line of the Gaussian random-bond Ising model.
The figure also contains a linear fit to the large size data
and the value $\sigma=1/1.02177$ predicted by duality arguments
\cite{NN}.}
\end{figure}

Figures with scaling collapses of $d_L$ were presented in
\cite{HPP,HJPP} with $\nu = 1.33$ and in \cite{MC} with $\nu =
1.50$. While they verify that $d_L$ obeys indeed the scaling form
(\ref{dScal}), the fact that reasonable collapses can be obtained for
different values shows that such a collapse is not a good criterion
for determining $\nu$.

\begin{figure}
\begin{center}
\epsfxsize=400pt{\epsffile{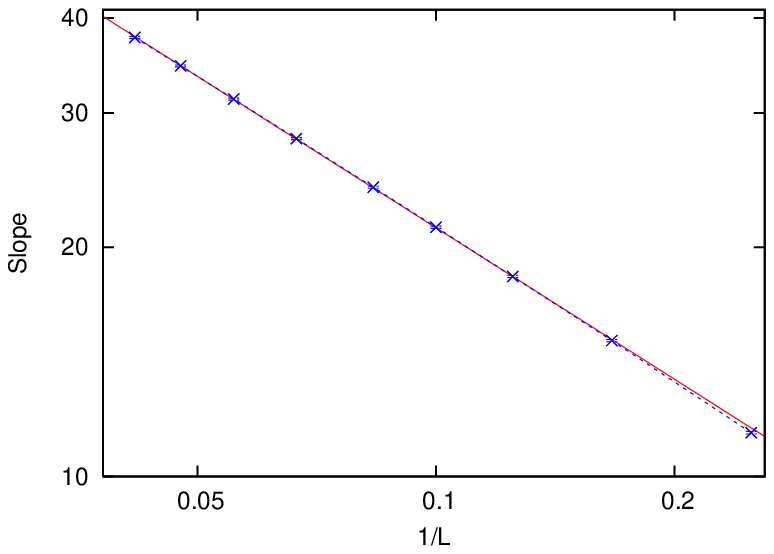}}
\caption{
Slope of the domain-wall free energy
of the Gaussian random-bond Ising model as a function of $1/L$.
\label{figSlopesg}
}
\end{center}
\end{figure}

\begin{figure}[t]
\begin{center}
\epsfxsize=400pt{\epsffile{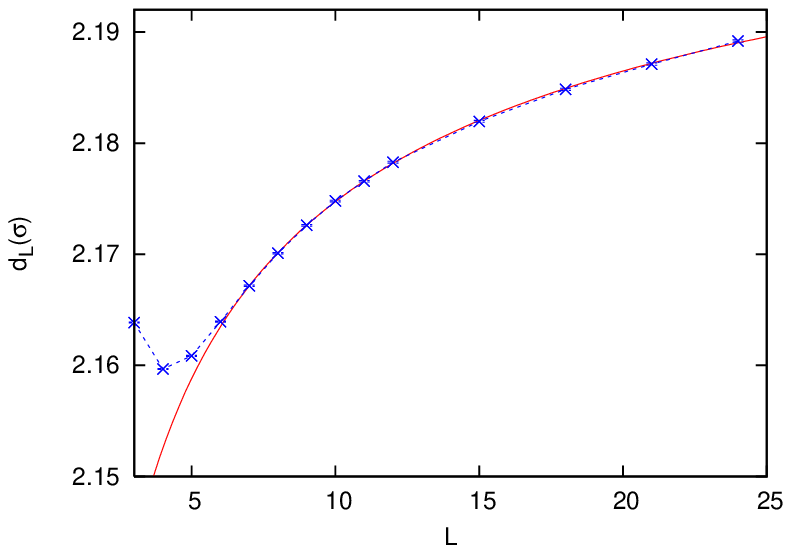}}
\caption{Domain-wall free energy
of the Gaussian random-bond Ising model at $\sigma_c=0.97945$ as a function of
$L$. We also show a plot of the fit to the form (\ref{eqmargin}) with
$L \geq 7$.
\label{figmargin}
}
\end{center}
\end{figure}

We now present the same domain-wall free energy analysis for the
Gaussian distribution of disorder. Again, $f_L^{(p)}$ and $f_L^{(a)}$
are computed with the transfer matrix technique for small
sizes. For larger sizes ($L=15,18,21,24$), we employed a different
algorithm, developed by Merz and Chalker, which uses a mapping to a
network model, see \cite{MC} for details. For each point we
averaged over $10~000$ samples of size $L \times 10^6$.
In Fig.~\ref{crossg}, we show a plot of the crossing point $\sigma(L_1,L_2)$
between $d_{L_1}(\sigma)$ and $d_{L_2}(\sigma)$.  Contrary to the $\pm
J$ case where the crossing was converging quickly, here we see that
the situation is much more complex. For the smaller sizes with
$L_2=L_1+2$ and $2/(L_1+L_2) \geq 0.15$, it apparently converges to
$\sigma \simeq 0.9815$. For larger sizes, we see a different
behaviour. We explain this change of behaviour by the existence of an
irrelevant operator. If such an operator exists, we expect
that close to the critical point, the domain-wall free energy can be
expanded as \cite{MM,MC,Sor}:
\beq
\label{eqmargin}
d_L(\sigma) = a + b \, (\sigma-\sigma_c) \, L^{1/\nu} + c \, L^{-x} \; .  
\eeq 
Here we have neglected a further contribution coming from the thermal
exponent $\nu_T$ (see \cite{MC}) which is sub-leading compared to the
$(\sigma-\sigma_c)$ term. The last contribution corresponds to an
irrelevant operator, of dimension $2 +x$, associated to the Nishimori
point. In principle it should also be present in the scaling of the
Nishimori point with the $\pm J$ distribution although much bigger
sizes may be necessary for its observation. If we take only the points
with $2/(L_1+L_2) < 0.1$ then the data in Fig.\ \ref{crossg} seems to
converge linearly to a point that we determine by a fit to be
\begin{equation}
\sigma_c = 0.97945 \pm 0.00004 \, .
\label{sigmaCritVal}
\end{equation}
In Fig.\ \ref{crossg} we also show the plot corresponding to this
linear fit (for the larger sizes) as well as the value
$\sigma=1/1.02177=0.978694$ predicted by duality arguments \cite{NN}.
While the two values are close,
they are still not compatible from our analysis. If one changes the
parameters of the linear fit, allowing for smaller size data to be
taken in account, the smallest possible value that we get is
$\sigma=0.97939 \pm 0.00002$ which is still not compatible with the
duality argument value. The only other numerical estimate of
$\sigma_c$ that we are aware of is the one of McMillan \cite{MM},
$\sigma_c \simeq 0.93-0.97$. This result is not expected to be highly
accurate since it is obtained on small systems of linear sizes up to
$L=8$.

Next we turn to the determination of $\nu$. Again, we compute it
by looking at the slope of the domain-wall free energy close to the
critical point. In Fig.\ \ref{figSlopesg}, we show the slope as a
function of the size. The slope is determined in the following way:
for each size, we computed $d_L(\sigma)$ for $\sigma=0.980$ and
$\sigma=0.982$. (For the sizes that we considered, the effective
critical value of $\sigma_c(L)$, determined as the crossing of the
domain-wall free energies, is always in this range, see Fig.\
\ref{crossg}). Thus the slope is determined directly from the
difference between these two values for each size. We also show in
this figure the best fit to the form $\simeq L^{1/\nu}$.
This fits perfectly the data for $L > 4$. We
obtain $\nu=1.52(3)$ which is very close to the result for the $\pm J$
disorder case.  Note that using a transfer matrix method McMillan
\cite{MM} obtained $\nu  \approx 1.58$ for the
Gaussian random-bond Ising model.

In Fig.\ \ref{figmargin}, we plot $d_L(\sigma)$ for the critical value
$\sigma_c=0.97945$ determined previously. One expects then that the
non-constant part comes entirely from the leading irrelevant
operator. A fit to the form (\ref{eqmargin}), also shown in
Fig.~\ref{figmargin}, gives excellent results if we remove the data
with $L\leq 6$. We obtain an exponent $x = 0.43(3)$.  Thus we predict
the existence of an irrelevant operator of dimension $2.43(3)$. We
observed previously that $\sigma_c(L)$ converges linearly (in $1/L$)
towards $\sigma_c(L\rightarrow +\infty)$, see Fig.\ \ref{crossg} . A
simple calculation starting from eq.(\ref{eqmargin}) shows that the
correction is of order $L^{-(1/\nu +x)}$ and the linear correction
corresponds to $x \simeq 0.33$. This is rather close to the
numerical result, the difference being easily explained by taking
into account further irrelevant operators. 

\section{Free energy and central charge}

\label{secC}

In this section, we use the free energy to
deduce the central charge. We obtain results for both the $\pm J$ and the
Gaussian distribution of disorder and compare these results to
check universality. 

As a byproduct of the measurements of the domain-wall free energy
presented in the previous section, one also obtains the free energy
$f_L^{(p)}$.  We will first discuss the $\pm J$ disorder case.
Averaging has been performed over sufficiently many samples to obtain
statistical errors of typically $\delta f_L^{(p)} < 4 \cdot 10^{-6}$.
Note that due to the aforementioned round-off error of one bond, $p$
has an error of about $10^{-7}$ on the strip sizes considered which is
not substantially below $\delta f_L^{(p)}$.

\begin{figure}[ht]
\begin{center}
%\leavevmode
\epsfxsize=400pt{\epsffile{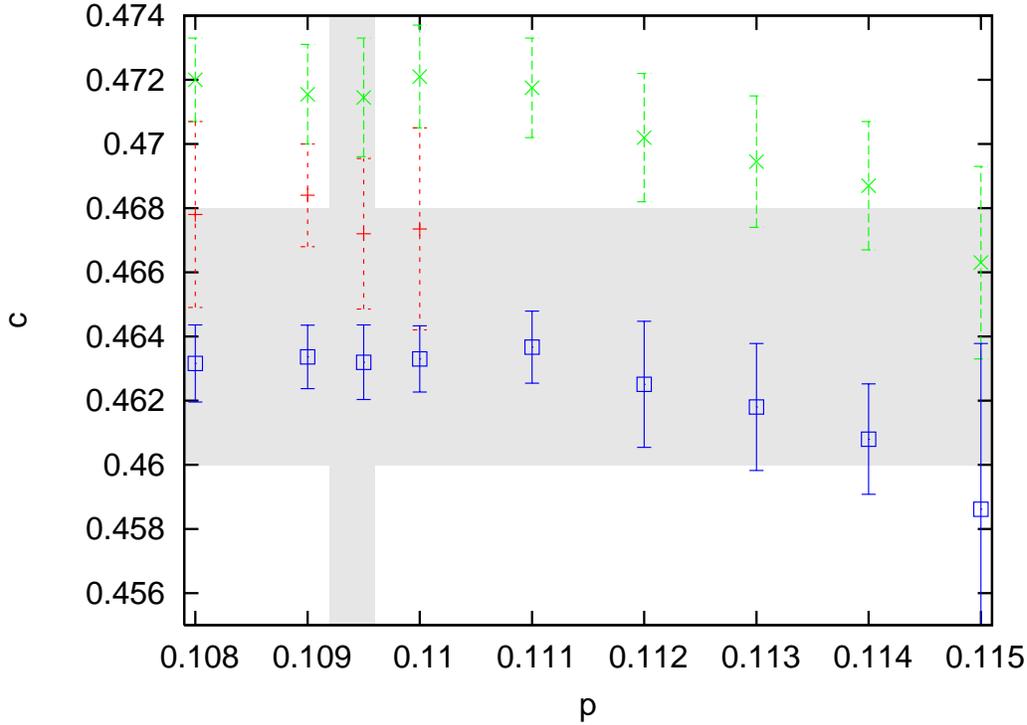}}
\end{center}
\caption{
Estimates for the central charge $c$ along the $\pm J$ Nishimori line
(\ref{pmJnis}) in the
vicinity of the Nishimori point obtained from (\ref{cEst2}) for
$5 \le L \le 9$ (`$\times$'), $6 \le L \le 12$ (`$+$')
or (\ref{cEst4}) with $4 \le L_0 \le 6$ (boxes). The
grey shaded areas denote our final confidence intervals for $p_c$
(vertical) and the central charge (horizontal).
\label{figC}
}
\end{figure}

The (effective) central charge $c$ can be estimated\footnote{For a
non-unitary theory what we call `$c$' is in fact
$c_{\rm eff} = c - 12 \, \delta_{\min}$
where $c$ is the central term in the operator
product expansion of $T$ with itself, and $\delta_{\min}$ the smallest
dimension contributing to the free energy. See section \ref{secCexp}
for further details.} from this data since it appears as the universal
coefficient of the first finite-size correction \cite{central}
\beq
f_L^{(p)} = f_{\infty}^{(p)} + {c \, \pi \over 6 L^2} + \ldots
\label{cEst2}
\eeq
Note that the leading term $f_{\infty}^{(p)}$ is not universal.
The universal $1/L^2$ term originates from the
energy-momentum tensor $T$
which has scaling dimension 2. The next correction is expected to arise
from $T^2$
and should therefore give rise to a term of the form $L^{-4}$
\beq
f_L^{(p)} = f_{\infty}^{(p)} + {c \, \pi \over 6 L^2} + {d \over L^4} + \ldots
\label{cEst4}
\eeq
Estimates for $c$ are obtained by fitting the free energy data in
intervals $L_0 \le L \le L_{\rm max}$ to (\ref{cEst2}) or
(\ref{cEst4}). We have considered only intervals with $L_0 \le L_{\rm
max}-3$ (at least four-point fits).  For a given form of the fit and
interval of lattice sizes, we have chosen $c$ to lie in the center of
the resulting fits and adjusted the error to include all fits.
The estimates for $c$ obtained in this manner are shown in Fig.\
\ref{figC}. Fits to (\ref{cEst2}) for lattice sizes $5 \le L \le 9$
are denoted by diagonal crosses and those for $6 \le L \le 12$
by plusses. Fits to (\ref{cEst4}) were performed for $4 \le L_0
\le 6$ with $L_{\rm max}$ kept fixed at the largest available system
size. The corresponding estimates are denoted by boxes.

One notices that the estimates in Fig.\ \ref{figC} are almost
independent of $p$ in the range considered.  On the one hand, this
means that the estimates for $c$ are stable. On the other hand, this
implies that estimates for the central charge are not a good tool for
locating the critical point precisely for the $\pm J$ distribution of
disorder (the situation will be different for the Gaussian
distribution, see below).  Furthermore, one observes a slight trend of
the fits (\ref{cEst2}) obtained with only a $L^{-2}$ term to shift to
smaller values as the range of system sizes considered is increased
while the fits (\ref{cEst4}) with a $L^{-4}$ term included converge
rapidly with system size. Therefore, we quote as a final estimate with
a generous error bar \cite{HPP}
\beq
c = 0.464 \pm 0.004 \, ,
\label{cVal}
\eeq
which is shown by the grey shaded horizontal bar (the estimate
(\ref{pcrit1}) for the location of the critical point $p_c$ is shown
by the grey shaded vertical bar).  In any case, all estimates obtained
from (\ref{cEst2}) with $L_{\rm max} = 12$ and $L_0 \ge 6$ satisfy $c
< 0.469$.  Furthermore, also consideration of other possible
finite-size corrections to (\ref{cEst2}) yields results which are
consistent with (\ref{cVal}). Thus, we can safely exclude the value
$c= {5 \sqrt{3} \ln{2} \over 4 \pi} \approx 0.47769$ for percolation
in the Ising model \cite{JC} even if the absolute difference is not
big.

\begin{figure}
\begin{center}
%\leavevmode
\epsfxsize=400pt{\epsffile{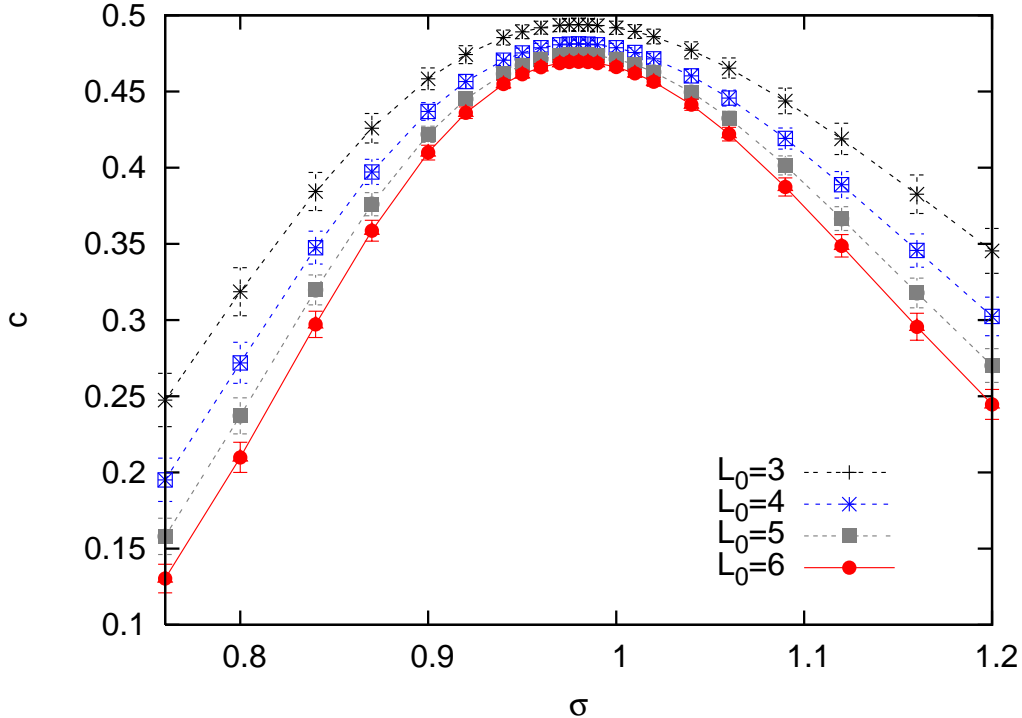}}
\end{center}
\protect\caption[2]{\label{cgaus} Central charge vs.\ $\sigma$ along
the Nishimori line for the Gaussian distribution of disorder, with $L_{\rm max}=8$.}
\end{figure}

We now turn to the Gaussian distribution of disorder. In this case, we
will argue that the central charge, along the Nishimori
line, gives a practical way of locating the multicritical point as the
point where the estimates of the central charge have a maximum.
We explained above that,
for the $\pm J$ distribution, the estimate of the central charge was
nearly independent of $p$ making it difficult to locate the
maximum. For the Gaussian distribution, the parameter $p$ is
replaced with the variance of the distribution $\sigma$, see
eq.~(\ref{gaussdis}). Thus, the situation is different since it is
possible to perform the simulations with the same disorder
configuration for each variance $\sigma$. More precisely, one defines
a bond $J$ in the following way: $J = J_0 + \sigma j$, $j$ being the
random part obtained from a random number generator which we choose to be
the same for different $\sigma$. As a consequence, most of the
measurements will be correlated and in particular this will be true
for the free energy and the deduced central charge (as was also the
case for the domain-wall free energy measured in the previous section).

\begin{figure}
\begin{center}
\epsfxsize=400pt{\epsffile{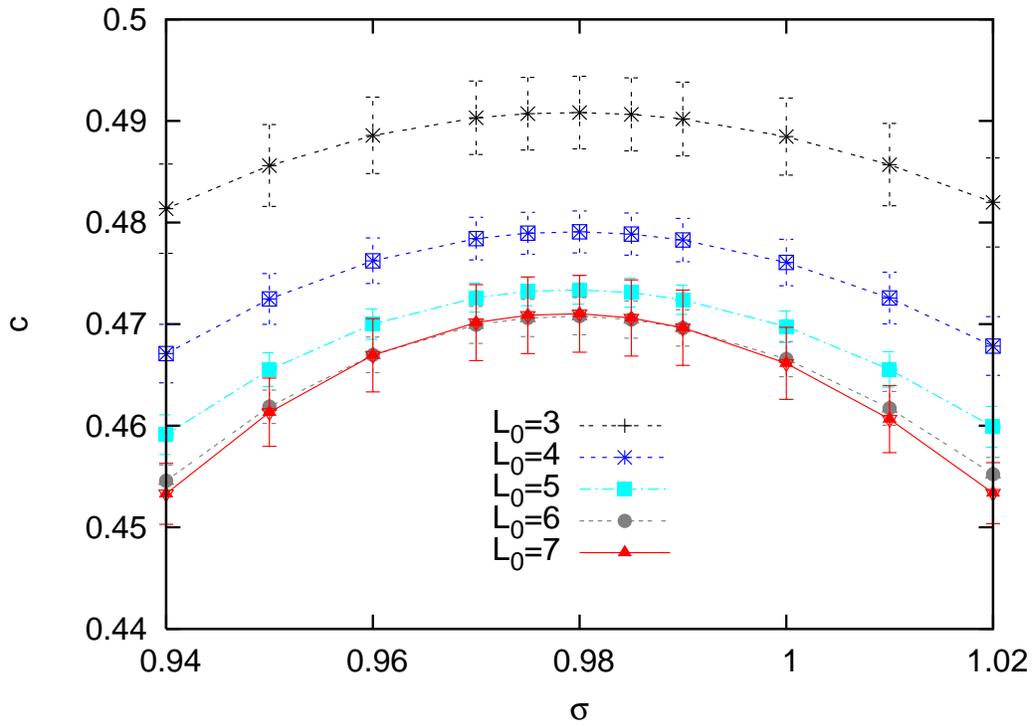}}
\end{center}
\protect\caption[2]{\label{cgausc} Central charge vs.\ $\sigma$ along
the Nishimori line for the Gaussian distribution of disorder close to the
maximum $\sigma_c \simeq 0.98$ and with $L_{\rm max}=10$.}
\end{figure}

\begin{figure}[t]
\begin{center}
\epsfxsize=400pt{\epsffile{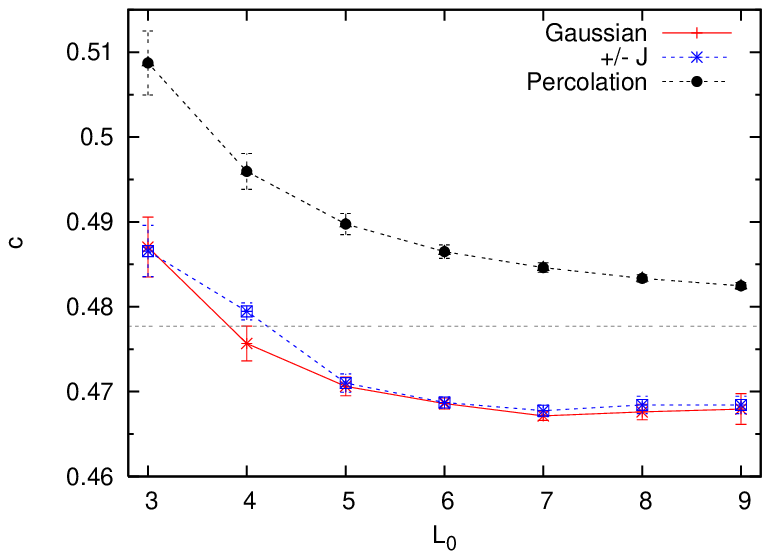}}
\end{center}
\protect\caption[2]{\label{ccc} Central charge at the multicritical
fixed point on the Nishimori line for the $\pm J$ and Gaussian
distributions of disorder. We also show for comparison finite-size
estimates of the central charge for percolation. The horizontal line
shows the infinite system limit $c= {5 \sqrt{3} \ln{2} \over 4 \pi}$
for percolation on Ising clusters \cite{JC}.
}
\end{figure}

The central charge is shown in Fig.\ \ref{cgaus} where we present the
effective central charge obtained by a fit to eq.\ (\ref{cEst2}) using
free energies in the range $[L_0,L_{\rm max}=8]$ for increasing
$L_0$. For the Gaussian distribution, we cannot suppress disorder
fluctuations by fixing the number of bonds with a certain value
globally.  As a consequence, the data shown in Fig.~\ref{cgaus} which
has been extracted from runs of the same size as for the $\pm J$
disorder (10~000 samples of strips of sizes $L\times 10^6$) has much
bigger error bars. Still these larger error bars are partially
compensated by the fact that the measurements are strongly
sample-to-sample correlated, and thus the maximum is easy to identify:
we have a stable maximum at $\sigma \simeq 0.98$. In Fig.\
\ref{cgausc}, we show the measured central charge close to the maximum
with $L_{\rm max}=10$. An estimate of the maximum gives for all the
ranges of $L$ employed a value of $\sigma_c \simeq 0.9805 \pm 0.0005$,
with no measurable change of the central charge compared to $\sigma =
0.98$. For the size that we consider here, the agreement is very good
with the $\sigma_c(L)$ obtained with the domain-wall measurements, see
Fig.\ \ref{crossg}. Additional runs were performed for the value
$\sigma = 0.98$, in particular for larger sizes, up to $L=12$. For
$\sigma=0.98$ the average is performed over at least 20~000 samples,
each sample being a strip of size $L \times 10^7$. We will employ in
the following this data to compute the central charge that we will
compare to the $\pm J$ disorder case. In Fig.~\ref{ccc}, we compare
the effective central charges obtained for the Gaussian and the $\pm
J$ distributions of disorder. For the $\pm J$ distribution, we
employed the data for $p=0.1095$. For comparison, we also show the
central charge obtained for percolation on an Ising model. In all
these cases, we show the central charge obtained by a fit to
(\ref{cEst2}) for increasing $L_0$ while keeping the maximum size
fixed to $L_{\rm max}=12$.  The agreement between the two types of
disorder is excellent. We see that the central charges converge to
$\simeq 0.465$ at the multicritical fixed point on the Nishimori line
for both cases of disorder. We thus have strong evidence of the
universality of the multicritical fixed point. Moreover, the
asymptotic value is clearly different from the value of percolation,
either the numerical one on a finite system or the infinite system
limit also shown in Fig.~\ref{ccc}.

\section{Spin-spin correlation functions}

\label{secCor}

In the previous sections, we have already seen that the universality class
of the multicritical point on the Nishimori line, for the two types of
disorder considered, is different from the one of
percolation. Still the central charge is very close and it is then
useful to find further measurements to confirm this
result. In this section, we present numerical results for the
spin-spin correlation functions on the Nishimori line close to
the Nishimori point.

A general correlation function $C_n$ between
two points $\vec{r}_1$ and $\vec{r}_2$
has the following power-law form with exponent $2\,x_n$
in a two-dimensional conformal field theory on the infinite plane:
\beq
C_n\left(\vec{r}_1, \vec{r}_2 \right) \propto
\frac{1}{\left| \vec{r}_1 - \vec{r}_2\right|^{2\,x_n}} \, .
\label{corInfPlane}
\eeq
In the transfer matrix computations we use long strips
with periodic boundary conditions along the short directions which
one can also interpret as a cylinder. Therefore, we
consider the correlation function $C_n$ on the infinite
cylinder of circumference $L$ with coordinates % $u, v$ with
$u\in [1,L]$ and $v\in ]-\infty,+\infty[$. Applying a conformal mapping
to (\ref{corInfPlane}), one infers the following behaviour on the 
cylinder (see, e.g., chapter 11.2 of \cite{Cardy})
\beq
C_n\left((u_1,v_1) , (u_2,v_2) \right) \propto 
{({2 \pi \over L})^{2 x_n}  \over 
\left\{2 \cosh(2\pi (v_1-v_2)/L) -2 \cos(2 \pi (u_1-u_2)/ L)\right\}^{x_n} } \, .
\label{fitsscf}
\eeq
%%
%\beq
%[\langle S(u_1,v_1) S(u_2,v_2)\rangle] \propto 
%{({2 \pi \over L})^{2 x_n}  \over 
%\left[2 \cosh(2\pi (v_1-v_2)/L) -2 \cos(2 \pi (u_1-u_2)/ L)\right]^{x_n} } \, .
%\label{fitsscf}
%\eeq
%
%Here $[\cdots]$ stands for the average over the disorder. 
There are two particular cases that we will consider in the following: 

\begin{itemize}
\item $v_1 = v_2$, $u=u_1-u_2$. The correlation function is measured
for two points across the strip, separated by a distance $u$. The
correlation function takes the form: 
\beq
%[\langle S(0) \, S(u)\rangle]
C_n(0,u) \propto
    \left(\sin({\pi u\over L}) L\right)^{- 2 \, x_n} \; .
\label{fitcf1}
\eeq
One can then extract the exponent $2\,x_n$ by a direct fit of the
measured correlation function, after averaging over the disorder, to
this form. One disadvantage of this method is that we can use only
points separated by distances $u \leq L/2$.

%\item If we consider the special case $u=L/2$ with again $v_1 = v_2$,
%the correlation function simplifies further and takes the form 
%%
%\beq
%%[\langle S(0) S(L/2)\rangle]
%_n(0,L/2) \propto  L^{- 2\,x_n} \; ,
%\label{fitcf2}
%\eeq
%which allows a direct determination of the exponent $2\,x_n$. 

\item % A third case is
When one chooses $u_1=u_2$, $v=v_1 - v_2$ the
correlation function takes the following form:
\beq
%[\langle S(0) S(y)\rangle]
C_n(0,v) \propto
    \left(\sinh({\pi v \over L}) L\right)^{- 2 x_n} \; .
\label{fitcf3}
\eeq
Thus we consider two points separated along the strip with the same
position across the strip, and it is possible to
access long distances.
% make fits
%of numerical data with only points far apart, thus reducing the effect
%of finite sizes.

\end{itemize}

We will apply these two forms to the $n$th moments of the
spin-spin correlation function
%
%\beq
$[\langle S(u_1,v_1) S(u_2,v_2)\rangle^n]$, % \, ,
%\label{defsscf}
%\eeq
where $[\cdots]$ stands for the average over disorder. 
The general result is then, for the two cases discussed above:
\begin{eqnarray}
[\langle S(0) S(x)\rangle^n] &\propto &
    \left(\sin({\pi x\over L}) L\right)^{- \eta_n} \; , \cr
%[\langle S(0) S(L/2)\rangle^n]  &\propto&   L^{- \eta_n} \; , \cr
[\langle S(0) S(y)\rangle^n]  &\propto& 
    \left(\sinh({\pi y \over L}) L\right)^{- \eta_n} \; .
\end{eqnarray}
Here, we have identified $\eta_n = 2 \,x_n$. All along the Nishimori line, the
moments of these correlation functions are equal two by two \cite{N}, thus we
have the general result that $\eta_{2k-1} = \eta_{2k}$. For a pure
system, one has $\eta_n = n \times \eta$.  On the other hand, in the
case of percolation over Ising clusters, it is easy to see that all
moments of the spin-spin correlation functions are equal (and not only two
by two). Then, if the Nishimori point would be in the percolation
universality class, the exponents for the correlation functions
should collapse to a unique value $\eta_n = \eta$ at
the critical point.

\begin{figure}
\begin{center}
%\leavevmode
\epsfxsize=400pt{\epsffile{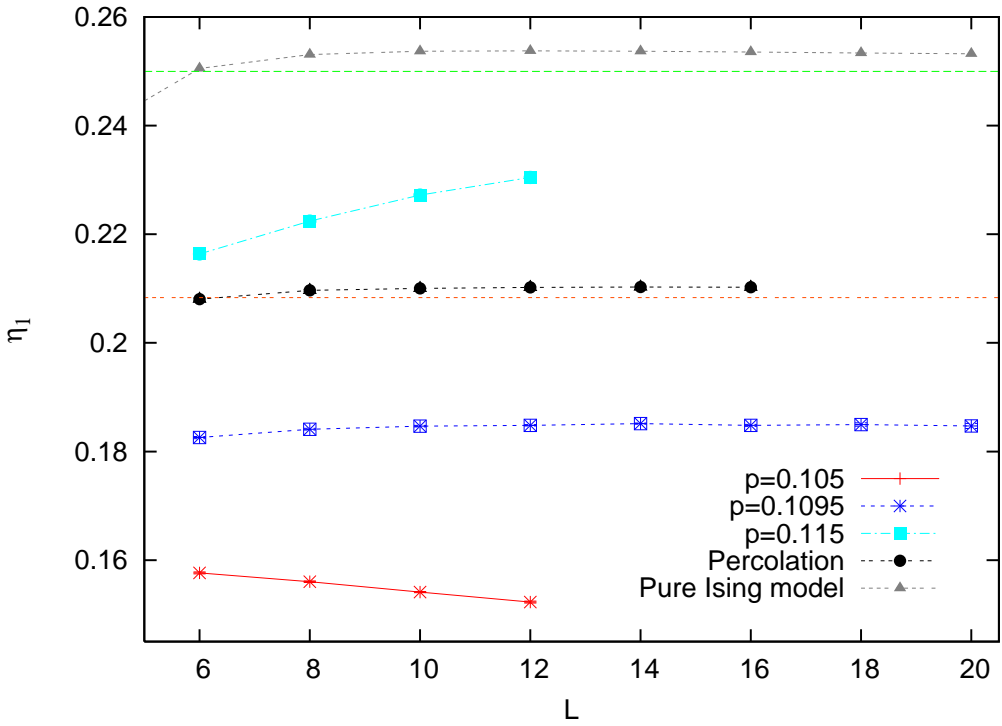}}
\end{center}
\protect\caption[2]{\label{ss1}Effective magnetic exponent $\eta_1$ for
the $\pm J$ RBIM with
$p=0.105$, $p=0.1095$, $p=0.115$, the percolation model and the pure Ising
model.} 
\end{figure}

As a first measurement for $\pm J$ disorder,
we show in Fig.\ \ref{ss1} the effective
magnetic exponent $\eta_1$ that we obtain from a fit to the form
eq.\ (\ref{fitcf1}). The effective magnetic exponent $\eta_1$
was computed for three points on the Nishimori line:
at $p=0.105$, which is well inside the ferromagnetic phase and indeed,
we see that the effective magnetic exponent tends to a small value;
at $p=0.115$ which is well in the paramagnetic phase, which is
confirmed by the fact that the effective magnetic exponent increases;
at $p=0.1095 \simeq p_c$ where the effective magnetic exponent seems
to converge to a value $\eta_1 \simeq 0.185$.
On this figure, we also show the corresponding magnetic exponent
obtained from simulations of the pure Ising model and also of the
percolation model, as well as the expected values for these two models
in the infinite limit ($\eta = {1\over 4}$ and ${5 \over 24}$,
respectively -- see, e.g., \cite{WK74} and \cite{StAh}, respectively).
% \footnote{The latter exponents can be obtained from the identification of
%critical percolation with the $q \to 1$ limit of the $q$ states Potts model, see for example
%\cite{StAh}.}.
These two measurements are presented in order to show
what type of correction we can expect in such a measurement of the
magnetic exponent. Indeed, we can see that this method gives rather
accurate measurements for large $L$ (for $L=12$, the deviation is
around $1\%$). From this, we can conclude that the value of $\eta_1$
for $p=0.1095$ is significantly distinct from the one of percolation.
This is an additional proof that the multicritical point on the Nishimori
line is not in the universality class of percolation.

\begin{figure}
\begin{center}
%\leavevmode
\epsfxsize=400pt{\epsffile{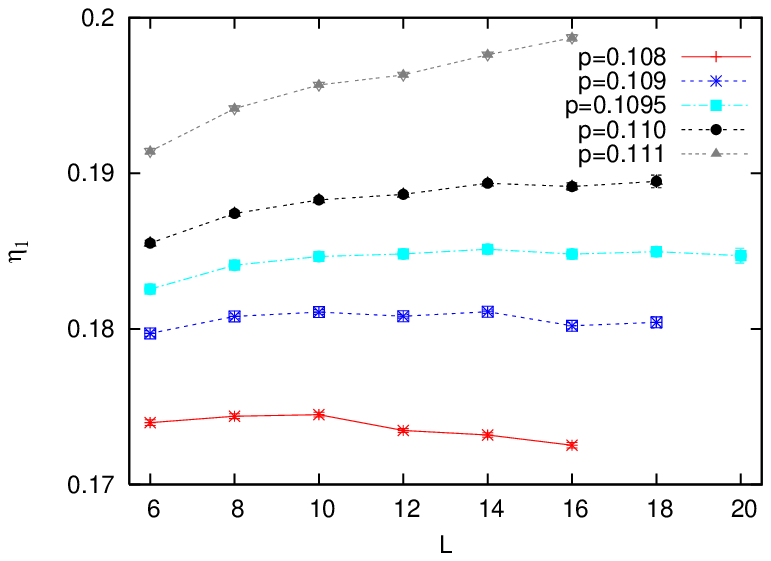}}
\end{center}
\protect\caption[2]{\label{ss1b}Effective magnetic exponent $\eta_1$ for
$p=0.108, \cdots, 0.111$, for the $\pm J$ disorder on the Nishimori line.} 
\end{figure}

Next, in Fig.\ \ref{ss1b}, we show the same quantity but only close to
the critical point. In this figure, one can clearly distinguish a change
of behaviour close to $p \simeq 0.1095$. For lower $p$, the exponent
decreases, which is expected since we are in the ferromagnetic phase.
On the contrary, for $p$ larger than $0.1095$, we observe that the
magnetic exponent increases, as expected since we are in the
paramagnetic phase.

\begin{figure}[ht]
\begin{center}
%\leavevmode
\epsfxsize=400pt{\epsffile{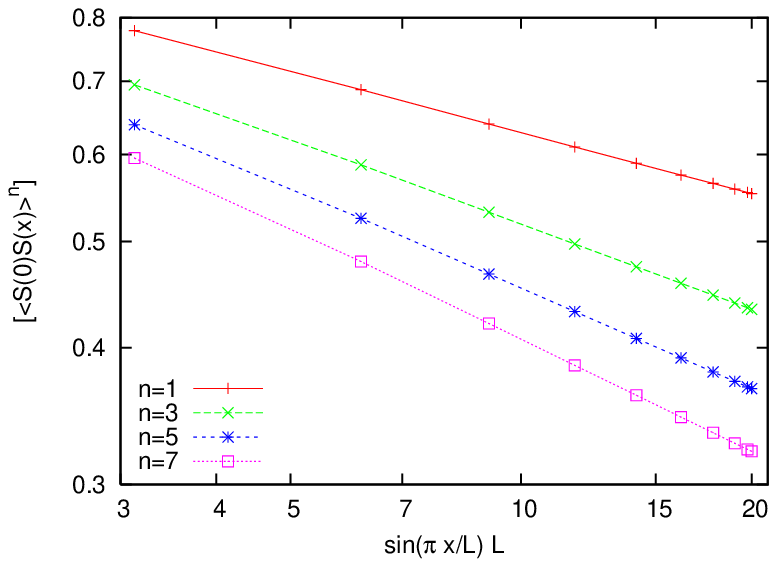}}
\end{center}
\smallskip
\caption{Moments of the spin-spin correlation function
for the $\pm J$ RBIM with
$p=0.1095$ and $L=20$. We only show the odd moments.
%: $n=1$ ($+$, red), $n=3$ ($\times$, green), $n=5$ ($*$, blue) and $n=7$
%(open boxes, pink).
Error bars are much smaller than the size of the symbols.
\label{ssFig}
}
\end{figure}

Fig.\ \ref{ssFig} shows the moments of the spin-spin correlation
function for $L=20$ and $p=0.1095$. There is a similar plot in
\cite{HPP}, but the present data set is completely independent. The
present data was obtained for a geometry which differs from
\cite{HPP}, namely on $4001 \times 20$ strips with a globally fixed
number of positive (negative) bonds while for that of \cite{HPP}, each
bond was assigned a value independently. Here we have discarded the
1000 initial (and final) iterations before taking 101 measurements of
the correlation functions every 20 iterations. The data in Fig.\
\ref{ssFig} was obtained by averaging over 7623 such strips, resulting
in statistical error bars which are much smaller than the size of the
symbols. Despite these differences, the present results for $[\langle
S(0) S(x)\rangle^n]$ agree with those of \cite{HPP} within error bars.

A direct fit on a doubly logarithmic scale of the correlation functions for
$p_c = 0.1095$ and $L=20$ to the form (\ref{fitcf1}) yields:
\bea
\eta_{1}&=&\eta_{2}= 0.1848 \pm 0.0003 \, , \nonumber \\
\eta_{3}&=&\eta_{4}= 0.2552 \pm 0.0009 \, , \nonumber \\
\eta_{5}&=&\eta_{6}= 0.3004 \pm 0.0013 \, , \nonumber \\
\eta_{7}&=&\eta_{8}= 0.3341 \pm 0.0016 \, .
\label{valExp20}
\eea
These estimates
are consistent with those of \cite{HPP,QSnew} within error bars.

\begin{figure}
\begin{center}
\epsfxsize=400pt{\epsffile{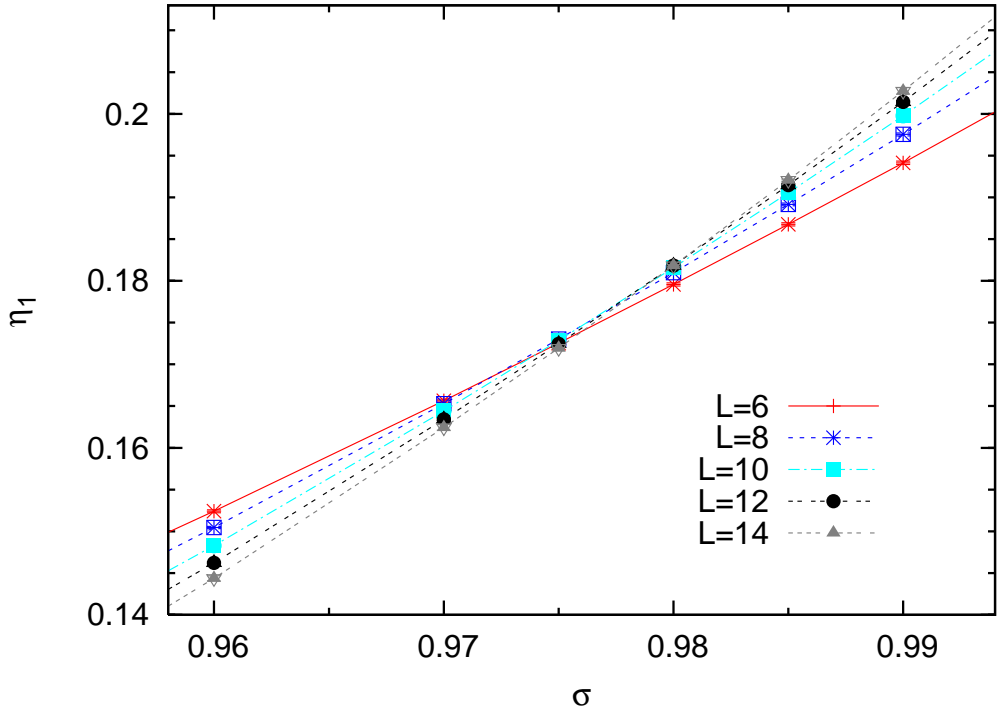}}
\end{center}
\protect\caption[2]{\label{mgaus2} Effective magnetic exponent $\eta_1$
vs.\ $L$ for the Gaussian distribution of disorder.}
\end{figure}

We now turn to the Gaussian distribution of disorder. In
Fig.\ \ref{mgaus2}, we present the exponent $\eta_1$ obtained from a
direct fit with eq.\ (\ref{fitcf1}) in function of $\sigma$. For that
case, we have data for size up to $L=14$ and for each value of $L$ and
$\sigma$ we average over 10~000 samples of geometry $L \times 200\,L$.
We clearly see a crossing of the curves close to $\sigma = 0.98$,
in agreement with the previous results. At the
crossing point, we have $\eta_1 \simeq 0.18$, which is very close to
the corresponding value for the $\pm J$ distribution of disorder.

Next, we present the exponents obtained, again for $L=14$, from a
direct fit on a doubly logarithmic scale to the correlation functions for 
$\sigma = 0.98$:
\bea
\eta_{1}&=&\eta_{2}= 0.1818 \pm 0.0002 \, , \nonumber \\
\eta_{3}&=&\eta_{4}= 0.2559 \pm 0.0002 \, , \nonumber \\
\eta_{5}&=&\eta_{6}= 0.3041 \pm 0.0002 \, , \nonumber \\
\eta_{7}&=&\eta_{8}= 0.3402 \pm 0.0002 \, .
\label{valExp14}
\eea
The value of these exponents is close to the ones for the $\pm J$
distribution of disorder. Note that there is still a small difference,
$\eta_1^{\pm J} \simeq 0.1848$ compared to $\eta_1^{\rm Gaussian} \simeq
0.1818$. This difference is due to the fact that the exponents are
obtained only close to the critical points. For the $\pm J$ disorder
case, the measurement is done at $p=0.1095$. From Fig.\ \ref{ss1b}, one
can read off that this will imply a change of order $0.002$ on $\eta_1$
if we take $p_c=0.1093$. Taking into account this correction, the
correspondence of $\eta_1$ is nearly perfect between the two types of
disorder, thus giving more support for the universality.

\begin{figure}
\begin{center}
\epsfxsize=400pt{\epsffile{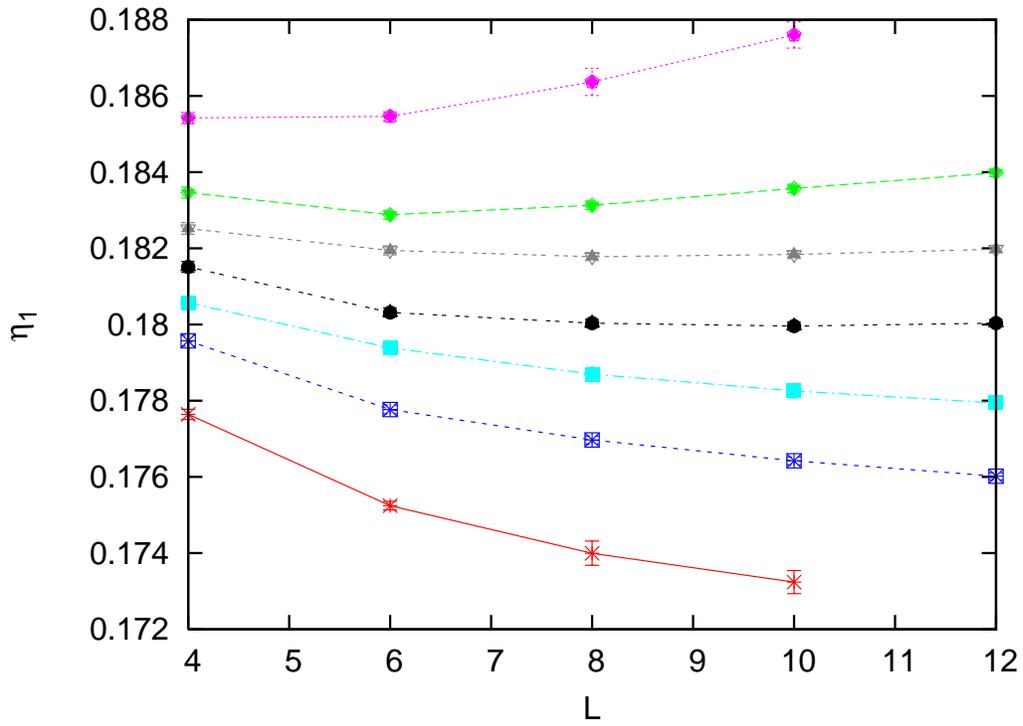}}
\end{center}
\protect\caption[2]{\label{mgaus4} Effective magnetic exponent
$\eta_1$ vs.\ $L$ for the Gaussian distribution of disorder with
$\sigma=0.976,0.978,0.979,0.980,0.981,0.982,0.984$ (from bottom to
top).}
\end{figure}
 
In Fig.\ \ref{mgaus4}, we show the exponent $\eta_1$ obtained by an
independent set of simulations on a geometry $L \times 100\,L$ and with
the measurement of the correlation function {\it along} the strip. We
simulated 10~000 samples for each size up to $L=12$ with this
geometry. The exponent $\eta_1$ is obtained from a fit with
eq.\ (\ref{fitcf3}). To perform the fit, we keep only the data for
correlation functions with two points at a distance $y$ such that $10
\leq y \leq 10\times L$. Thus one does not use the data for two
operators very close, contrary to what is done while fitting with eq.\
(\ref{fitcf1}) and we expect to reduce the finite-size corrections. In
Fig.~\ref{mgaus4}, one sees that for the largest $L$, one obtains a
constant exponent for $\sigma=0.98-0.981$ with $\eta_1 =
0.180-0.182$. Thus these results are in perfect agreement with the
previous measurements on a different geometry.

\begin{figure}
\begin{center}
%\leavevmode
\epsfxsize=400pt{\epsffile{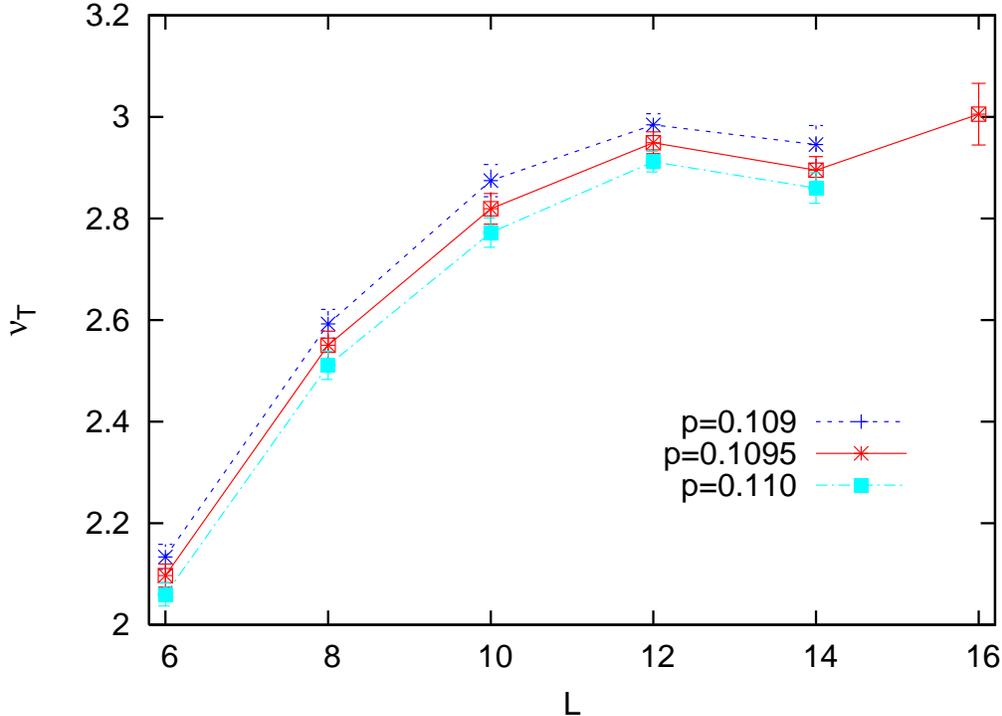}}
\end{center}
\protect\caption[2]{\label{mgaus5} Thermal exponent $\nu_T$
vs.\ $L$ for the $\pm J$ distribution of disorder.}
\end{figure}
 
Finally, in Fig.\ \ref{mgaus5}, we present results for the correlation
function of the energy operator $\epsilon_{ij}=J_{ij}S_i S_j$.
A thermal exponent $\nu_T$ is associated to
this operator, which corresponds to a perturbation in temperature,
via the relation
\beq 
\nu_T = {1\over (2 - x_T)} \; .  
\eeq 
This exponent replaces the exponent $x_n$ in eq.\ (\ref{fitcf1}) if we
replace the spin operator with the energy operator. In Fig.\
\ref{mgaus5}, $\nu_T$ is shown for $p \simeq p_c$ and for increasing
strip widths $L$. Finite-size corrections are very strong
for $\nu_T$. We obtain a value of $\nu_T \simeq 3$ for $L=14,\ldots,16$,
but it is not clear if we have reached a large enough size. With a
different method, Merz and Chalker \cite{MC} obtained a value of
$\nu_T = 4 \pm 0.5$ by measurements on system sizes up to $L=32$. We
believe that the extrapolation of our results for larger sizes is
compatible with this result.

\subsection{Central charge and magnetic exponents}

\label{secCexp}

We now return to the central charge. We mentioned at the beginning
of section \ref{secC} that the measured central charge 
is an effective quantity  which can be affected by operators with 
negative dimensions. Indeed, if one considers a torus, \ie\ a strip of
geometry $M \times L$ with periodic boundary conditions in both
directions, one has the relation \cite{cardyr}
\beq 
{Z \over Z_{\rm bulk}} = Q^{-c/12} \sum_{a,b} N_{a,b} Q^{\delta_a + n_b}
\eeq
with 
\beq
Q= e^{ -2 \pi {M \over L}} \; .
\eeq
Here, the index $a$ is associated to the primary operators $\phi_a$
which appear in the transfer matrix, while the index $n_b$ is non zero
(and positive) for the conformal descendants only. $N_{a,b}$ counts the 
multiplicity of the descendants of the
operator $\phi_a$ at the level $b$. 
Next, we consider the free energy per spin:
\beq
f(L,M) = {\ln{Z(M,L)} \over M L} = f_{\rm bulk} + {c \, \pi \over 6 L^2} +
{1\over M L } \ln{\left[ \sum_{a,b} N_{a,b} Q^{\delta_a + n_b} \right]} \; .
\label{eqLM}
\eeq
If all the operators which appear in the transfer matrix have a
non-negative dimension, then in the limit $M \gg L$, the last term in 
eq.~(\ref{eqLM}) can be dropped and one recovers eq.~(\ref{cEst2}).
On the contrary, if an operator with 
a negative dimension is present in the transfer matrix,  then for large $M$,
the last term in eq.~(\ref{eqLM}) will be
dominated by this operator. Let us call $\delta < 0$ the dimension
associated to the operator with the lowest dimension. Then
eq.~(\ref{eqLM}) will become, in the limit $M \gg L$:
\beq
f(L,M) = f_{\rm bulk} + {\pi \over 6 L^2}
 (c -12 \,\delta)  + \ldots
\eeq
Thus, if a negative dimension operator appears in the transfer matrix,
the measured central charge will be only an effective quantity
$c_{\rm eff}=c-12 \, \delta > c$. If no negative dimension operators appear
in the transfer matrix, then one has $\delta=0$ since the identity
operator with zero dimension is always present. 

\begin{figure}
\begin{center}
\epsfxsize=400pt{\epsffile{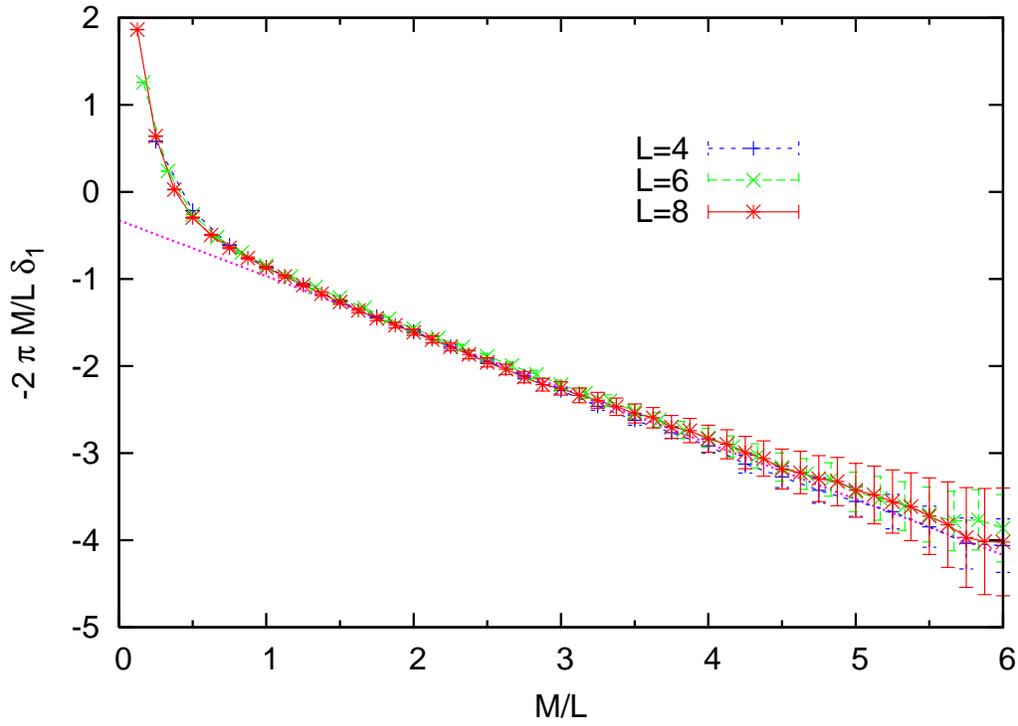}}
\end{center}
\protect\caption[2]{\label{reccc} Fit of the free energy for the
$\pm J$ RBIM with $p=0.1094$ to the form (\ref{estDelta1})
in the range $1.5 \leq M/L \leq 5$ for $L=4, 6$ and $8$.}
\end{figure}

It is known that negative exponents arise at the Nishimori point
in certain moments of the correlation functions of the disorder operator
\cite{MC2}. Still, these operators will not cause
any problems in the present context since the
disorder operator is not a local operator of our theory. 

We know of no direct way to determine if an operator with a negative dimension 
is present in the transfer matrix. Still, one can make the following simple test. 
In the pure Ising model, the lowest dimension corresponds to the magnetic operator. 
Since one expects that the magnetic exponent is present and we know that its
dimension is small, one can compare 
it with the first dimension which appears in eq.~(\ref{eqLM}). Assuming 
that there is no negative dimension operator in the transfer matrix, the last 
part of eq.~(\ref{eqLM}) takes the form
\beq
{1\over M L } \ln{\left( 1 + e^{(-2 \pi {M\over L} \delta_1)} + \ldots  \right)}
\eeq
with $\delta_1$ the lowest dimension. Thus, by computing 
\beq
\ln{\left( e^{ M L (f(L,M) - f(L,M\rightarrow +\infty))} -1 \right)} 
\simeq -2 \pi {M\over L} \delta_1 + \ldots  \; ,
\label{estDelta1}
\eeq
one can estimate directly $\delta_1$. In Fig.~\ref{reccc}, we show
this quantity for $L=4,6$ and $8$, as well as a fit to the data in the
range $1.5 \leq M/L \leq 5$. These bounds are selected by imposing a
good quality of the fit. The values extracted for $\delta_1$ are
$0.108(1),0.104(1),0.102(1)$ for $L=4,6,8$, respectively. These values
seem to converge to a value close to the one of the magnetic exponent
$\eta_1/2 \simeq 0.0925$ such that it is reasonable to identify it with
the magnetic operator. This suggests that the measured central
charge is indeed the real central charge. 

\section{Other measurements}

In this section, we present measurements of other quantities, namely
the Binder cumulant and the magnetic susceptibility. For each of
these quantities, we do not expect to improve precision, the
measurements are rather done in order to check the consistency
of the previous measurements.

\subsection{Binder cumulant}
We first present measurements of the magnetic Binder cumulant on the
Nishimori line, in order to perform an independent measurement of the
critical $p$ for the $\pm J$ RBIM.
The magnetic Binder cumulant is defined as follows
in terms of the moments of the magnetization $m$ \cite{Binder81,LB00}:
\beq
B(L)  =  {1\over 2} \left(3 - { \left[\langle m^4\rangle \right]\over
\left[\langle m^2\rangle\right]^2}\right) \; .
\eeq
We note that on the Nishimori line the magnetic Binder
cumulant is identical with the overlap Binder cumulant which is
usually employed in measurements for spin-glass models,
since we have an equality of the first and second moment of the
spin-spin correlation functions \cite{N}.

\begin{figure}
\begin{center}
\epsfxsize=400pt{\epsffile{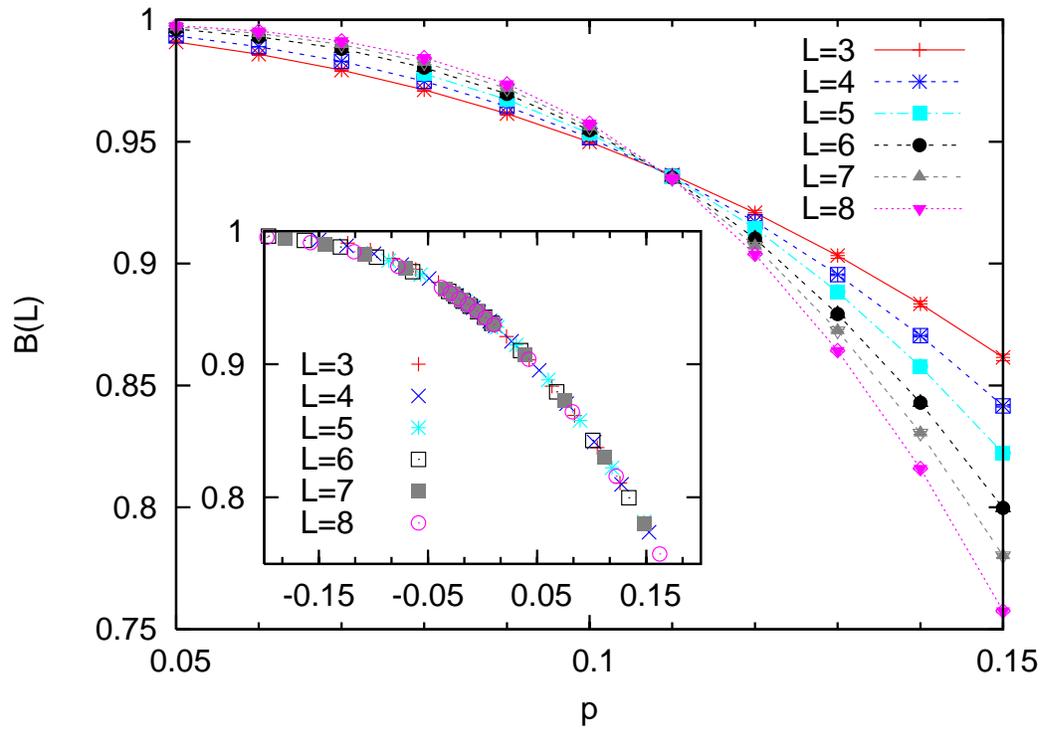}}
\end{center}
\protect\caption[2]{\label{b1} $B(L)$ vs.\ $p$ on the Nishimori line
for the $\pm J$ disorder case. The inset shows $B(L)$
vs.\ $(p-p_c) L^{1/\nu}$, with $p_c=0.1094$ and $\nu=3/2$.}
\end{figure}

Simulations are performed on square lattices with periodic boundary
conditions in both directions and with linear sizes in the range $L=3$
up to $L=8$. We employ the transfer matrix to compute the
partition function without and with a small magnetic field $h$ as
well as with $2h$. Then we extract the second and fourth moment of
the magnetization $m$ from the expansion
\beq 
Z(h) = Z(h=0) \left(1 + {h^2 \over 2} \, \langle m^2 \rangle
 + {h^4 \over 4!} \, \langle m^4\rangle + \cdots \right) \; , 
\eeq 
and a similar expansion of $Z(2h)$. The terms $\langle m \rangle$ and
$\langle m^3 \rangle$ do not appear in the expansion since they vanish
at $h=0$ on a finite system. We used a value of $h \simeq 0.01 / L^2
$ in our simulations. In order to reach a good precision, a large
number of samples had to be simulated, typically one million samples
for each size and value of disorder $p$. Fig.~\ref{b1} shows a
plot of the Binder cumulant versus $p$. In this figure, we observe
a crossing in the expected region, \ie\ $p \simeq 0.11$. Since
the number of samples that we have to simulate is huge, it is
difficult to reach sizes large enough to improve the previous estimate
of the critical point on the Nishimori line. Thus the measurements
that we show here should be regarded as a consistency test only.

The inset of Fig.~\ref{b1} shows a plot of the Binder cumulant versus the
rescaled variable, $(p-p_c)\,L^{1/\nu}$. Assuming the values that we
obtained in section \ref{secDW}, $p_c=0.1094$ and $\nu=3/2$, we see that
we have a good scaling behaviour of the Binder cumulant already for
small lattices.

\begin{figure}
\begin{center}
\epsfxsize=400pt{\epsffile{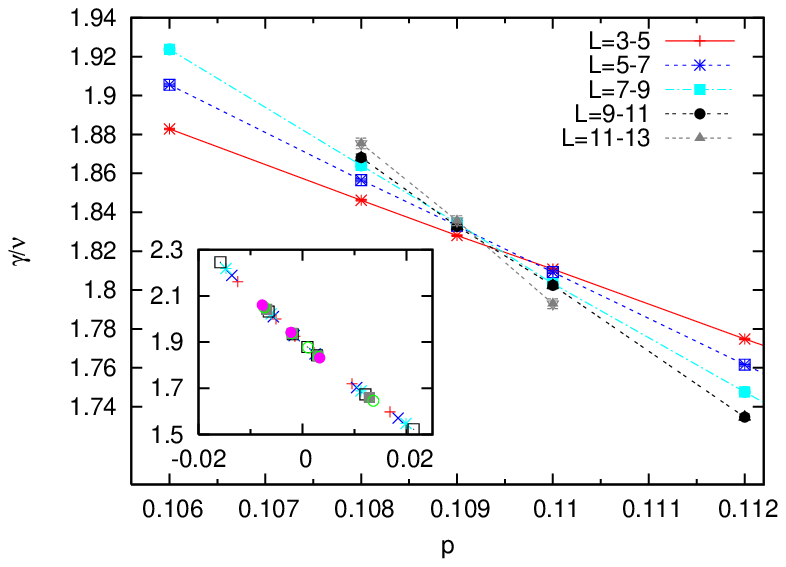}}
\end{center}
\protect\caption[2]{\label{sn5} Susceptibility exponent
$\gamma/\nu$ vs.\ $p$ on
the Nishimori line for the $\pm J$ disorder case. The inset
shows $\chi (L) L^{-\gamma/\nu}$ vs.\ $(p-p_c)L^{1/\nu}$, with
$p_c=0.1094$ and $\nu=3/2$.}
\end{figure}

\subsection{Susceptibility}

\label{secSusc}

A second quantity of interest is the susceptibility $\chi$, which we measure
using the transfer matrix. We compute the free energy, first without
any applied magnetic, and next with a small magnetic field.
$f_L(0)$ and $f_L(h)$ are determined with
the same realization of disorder in order to decrease the
fluctuations. The susceptibility is obtained by the following relation
\beq f_L(h)-f_L(0) 
\simeq (\beta h)^2 \chi (L) \; .
\eeq 
We need to choose a very small magnetic field $h$, typically $h \simeq
0.000001$, in order to ensure that we do not have a magnetic term.

We expect that the susceptibility is of the form $\chi (L) \simeq
L^{\gamma/\nu}$ at a critical point. In Fig.~\ref{sn5}, we
present the effective susceptibility exponent obtained by two-point
fits for data with odd sizes. For each of these points, we made
$10~000$ measurements on strips of sizes $L \times 10^5$. We see that the
crossing converges towards a value of $p$ close to $0.109$. Moreover,
we can see that the value of the susceptibility exponent seems to
converge to a value close to $\gamma / \nu \simeq 1.82$. Thus one can
also check that the hyperscaling relation ${\gamma/ \nu} + 2{\beta
/\nu} = d$ is satisfied with $2 {\beta / \nu} = \eta_1 \simeq 0.18$.

Finally, we can also use the measured values of the susceptibility to
perform a fit in the following form
\beq
\label{fchi}
\chi(L) \simeq L^{\gamma/\nu} a((p-p_c) L ^{1/\nu}) \; .
\eeq
Since this fit involves three parameters, we obtain a
large error bar on each of them. Keeping only the
reasonable fits, one obtains good collapses of the data in
the following range: $\gamma/\nu = 1.8 - 1.82, p_c=0.109 - 0.110, 1/\nu =
0.65 - 0.7$. In the inset of Fig.\ \ref{sn5} we present a plot of $\chi(L)
L^{-\gamma/\nu}$ vs.\ the rescaled variable $(p-p_c) L ^{1/\nu}$ with the
values $p_c = 0.1094$, $1/\nu = 2/3$ and $\gamma /\nu = 1.82$ with an
excellent collapse of the data.

\section{Dilution}

\begin{figure}
\begin{center}
%\leavevmode
\epsfxsize=400pt{\epsffile{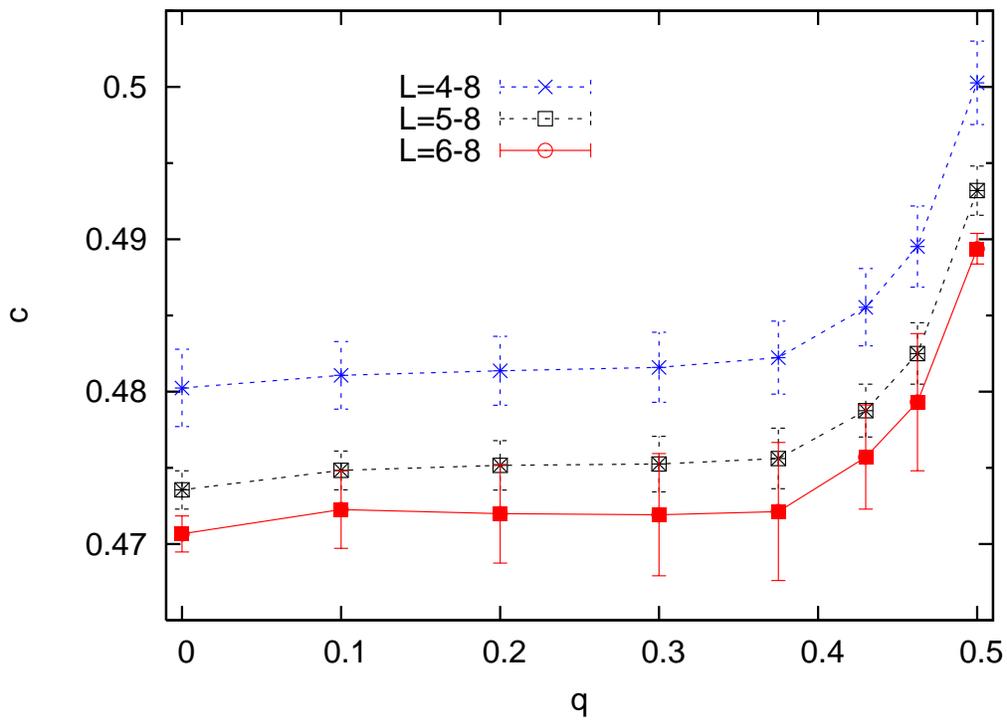}}
\end{center}
\protect\caption[2]{\label{cdil0} Central charge vs.\
dilution $q$ along the intersection of the $p-q-T$ Nishimori surface
with the paramagnetic-ferromagnetic critical surface for
the $\pm J$ disorder case.}
\end{figure}

In this section, we will consider the more general case of a binary
distribution with dilution. We denote by $q$ the amount of dilution
and by $p$ the amount of disorder (see section 2 for the definitions).
The Nishimori line is now replaced by a surface in the $T-p-q$ space.
For $p=0$, we expect only two critical points \cite{YeSt79}: an
attractive point for $q=0$ (no dilution) and a repulsive point for
$q=0.5$ which is a percolation fixed point (see figure \ref{pdbin})
and is on the Nishimori surface.  Another fixed point, also on the
Nishimori surface, is the fixed point on the Nishimori line determined
previously for $q=0$. In this section we want to study the flow
between these two fixed points on the Nishimori surface. In
particular, obtaining a clear flow between these two fixed points will
give further evidence that the multicritical point on the Nishimori
line ({\it i.e.}\ without dilution) is not in the same universality
class as percolation. Our measurements are carried out as follows: for
a fixed dilution $q$, we perform simulations for varying $p$ and $T$
in the Nishimori surface defined in eq.~(\ref{nissurface}) and look
for a maximum in the effective central charge. Next we check the value
of this maximum $c_{\rm eff}(q)\ {\rm vs.}\ q$ which is shown in Fig.\
\ref{cdil0}, for central charges obtained numerically with a fit to
the form eq.~(\ref{cEst2}) for the sizes $[4,5,6,7,8]$, $[5,6,7,8]$
and $[6,7,8]$. As expected, $c_{\rm eff}(q)$ varies monotonically
between $c_{\rm eff}(q=0)$ and $c_{\rm eff}(q=0.5)$ showing that there
is no additional fixed point.

\begin{figure}
\begin{center}
%\leavevmode
\epsfxsize=400pt{\epsffile{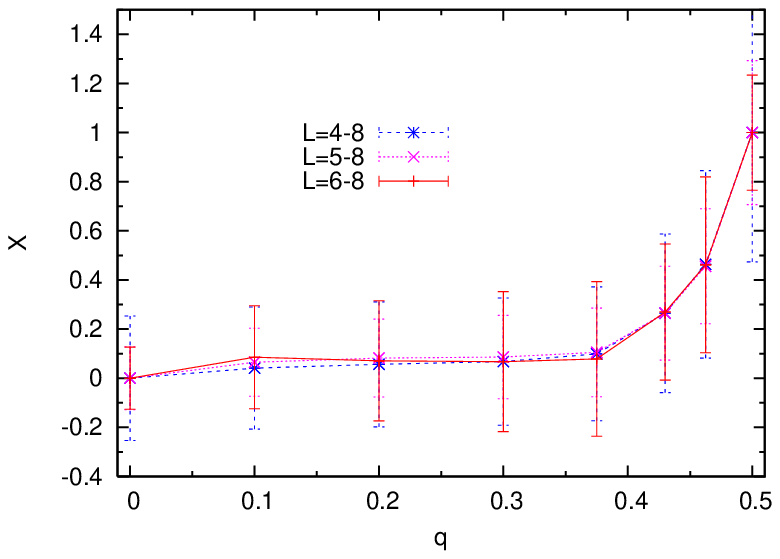}}
\end{center}
\protect\caption[2]{\label{cdil} $X$ vs.\
dilution $q$ along the intersection of the $p-q-T$ Nishimori surface
with the paramagnetic-ferromagnetic critical surface for
the $\pm J$ disorder case.}
\end{figure}

\begin{figure}[t]
\begin{center}
%\leavevmode
\epsfxsize=400pt{\epsffile{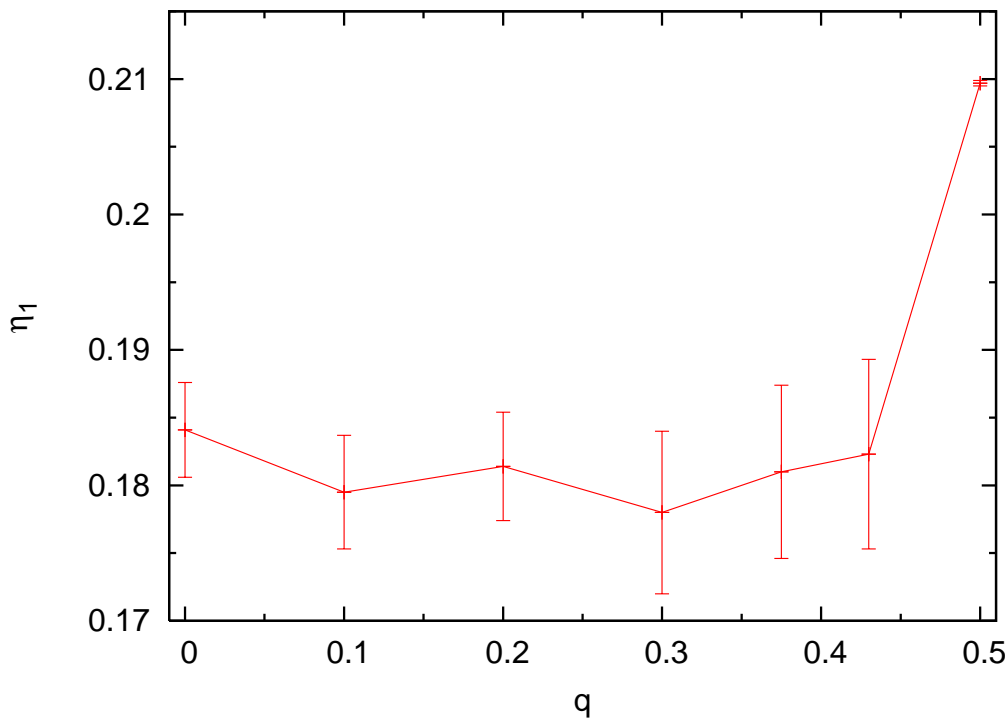}}
\end{center}
\protect\caption[2]{\label{cdilM} Magnetic exponent $\eta$ vs.\
dilution $q$ along the intersection of the $p-q-T$ Nishimori surface
with the paramagnetic-ferromagnetic critical surface for
the $\pm J$ disorder case.}
\end{figure}

In practice, since $c_{\rm eff}$ changes on a very small range
(between $0.47769$ for $q=0.5$ and $\simeq 0.464$ for $q=0$) with
strong finite-size corrections, it is more convenient to consider the
following quantity:
\beq
\label{Xdef}
X = {c_{\rm eff} - c_{\rm eff}(0) \over c_{\rm eff}(0.5)-c_{\rm
eff}(0)} \; ,
\eeq 
where for each range of sizes $L$ we consider the values obtained
with the same corrections for $c_{\rm eff}(0)$ and $c_{\rm
eff}(0.5)$. By construction, this quantity is equal to zero at $q=0$
(no dilution) and is equal to 1 at $q=0.5$ (the percolation case).
Fig.~\ref{cdil} shows the quantity $X$ as a function of dilution
$q$, again for the sizes $[4,5,6,7,8]$, $[5,6,7,8]$ and $[6,7,8]$.
Apart from the size of the errors, one does not observe significant
differences. In all cases, one immediately sees that the central charge
is dominated by the behaviour of the fixed point without dilution. If
one starts at the percolation point, a small decrease in $q$ will
appear as a jump towards the fixed point without dilution
($q=0$)\footnote{More precisely, critical properties at finite temperature
are expected to be controlled by the fixed point N' in Fig.~\ref{pdbin}
\cite{LG}. Fortunately, the crossover scale for the flow between N
and N' appears to be so small that this difference can be neglected
in practice.}. At $q\simeq 0.4$, $X$ is already indistinguishable
from zero, indicating that the central charge is the
same as the one for $q=0$. One expects that by increasing the sizes of
the data in the fit for determining $c_{\rm eff}$, one should observe
a crossover between the two fixed points $q=0.5$ and $q=0$ (such an
effect will be shown in the next section for the flow between the
Nishimori point and the fixed point of the pure Ising model). Here we
cannot really observe this effect since the crossover is too fast.

Finally, Fig.~\ref{cdilM} shows the magnetic exponent vs.\ the
dilution measured on long strips of width $L=8$. Here again, we
observe the same effect. For $q=0.5$, the magnetic exponent is known
exactly, it is $\eta = 5/24 \simeq 0.208333$ \cite{StAh}. As we decrease $q$,
the magnetic exponent jumps to a value compatible with the one of the
multicritical point on the Nishimori line $\eta \simeq 0.18$.

\section{Out of the Nishimori line}

In this section, we present some results off the Nishimori line. We
will consider two cases separately. First we investigate the line which
corresponds to the flow from the Nishimori point to the pure
Ising model. This line corresponds to the ferromagnetic-paramagnetic
transition line. We show in the next subsection that this line
can be determined from the maximum of the central charge when varying
temperature at fixed $p$. We show further that the flow on this line is
from the Nishimori point towards the pure Ising model fixed
point. Next we turn to the line which connects the
Nishimori point with a fixed point at zero temperature. An
important issue concerns the verticality or
re-entrance of this line. By arguing that the nature of the Ferro-Para
transition at the Nishimori point is of geometric origin, it was
suggested \cite{NK,ON} that the transition should take place at the
same concentration of impurities for any temperature below
the Nishimori point, implying verticality of the line. However, recent
numerical results for the $\pm J$ RBIM seem to advocate instead a
re-entrance of the paramagnetic phase (see Table~\ref{sumPc}). The other
important issue is the nature of the fixed point at zero temperature. It
was argued in light of previous numerical results that the
universality class of this point could be percolation \cite{SA,KaRi},
although there is no obvious reasoning supporting this conclusion in
contrast to the case of dilution.

\subsection{Flow from the Nishimori point to the pure Ising model}

First we study the line which separates the ferromagnetic
phase from the paramagnetic phase between the
Nishimori point and the pure Ising model. The phase boundary itself
has already been determined with good accuracy \cite{MC}. Here
we will use a computation of the central charge to follow this
line in a similar way to what was done in the case of dilution.
The motivation for studying this line is first to clearly
see in which direction we flow (we expect towards the pure Ising model
fixed point since this is a marginally attractive fixed point \cite{DDSL}).
Furthermore, we want to check that there is no
additional fixed point. In a similar study for the $3$-state Potts
model \cite{JP}, an additional fixed point, predicted by perturbation
theory \cite{Ludwig,DPP} was observed.

\begin{figure}
\begin{center}
\epsfxsize=400pt{\epsffile{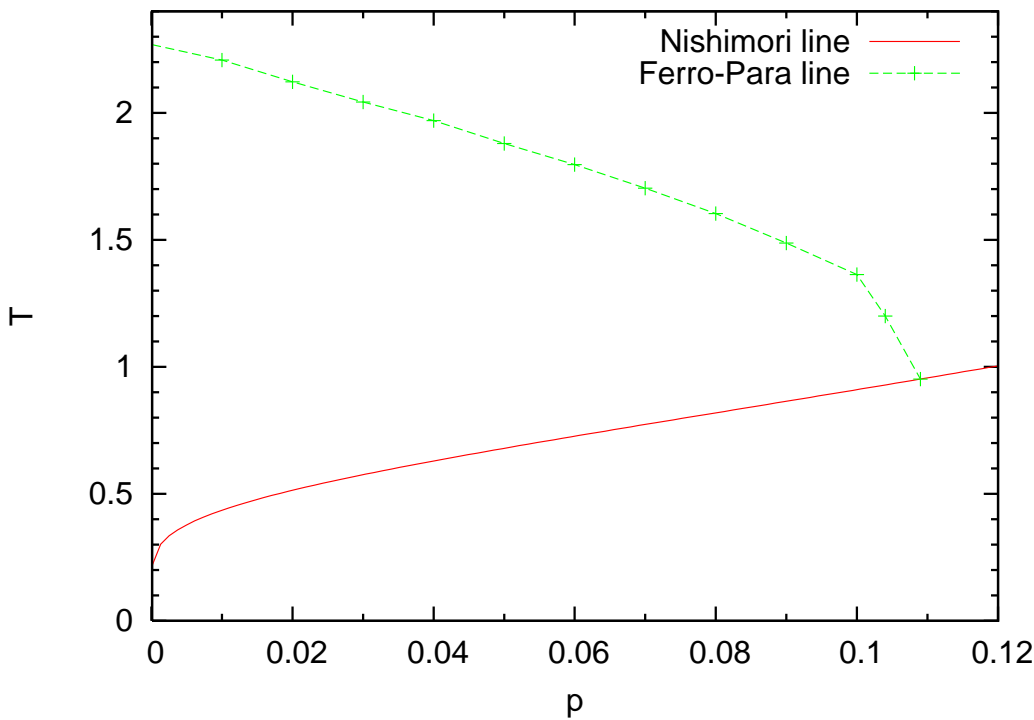}}
\end{center}
\protect\caption[2]{\label{phd} $T-p$ phase diagram for the
$J=\pm 1$ disorder case.}
\end{figure}

\begin{figure}
\begin{center}
\epsfxsize=400pt{\epsffile{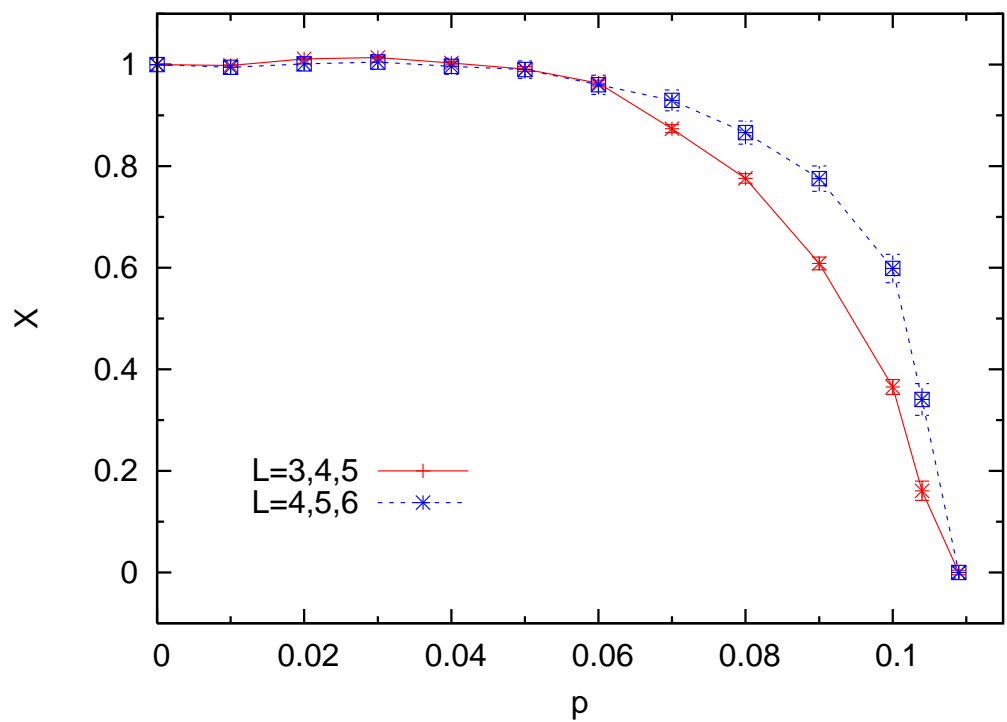}}
\end{center}
\protect\caption[2]{\label{ccfp} Variation of the effective central charge
as characterized by the quantity $X$ (see (\ref{label8.1}))
on the Ferro-Para line, for the $J=\pm 1$ disorder case.} 
\end{figure}

We start from the pure Ising model, with a small perturbation, say
$p=0.01$ and $T \simeq T_c$, $T_c$ being the critical temperature of the
pure Ising model. Next we vary $T$ and measure the central charge. For
small $p$, these measurements are very simple to perform since we have
only a weak disorder and moreover, measurements are performed at
constant $p$. It is then easy to determine a maximum since the same
configurations of disorder are employed for different $T$. Then we
iterate the process for larger $p$ and follow the ferro-para line
which is identified with the maximum of the central charge. The
measurement is more complicated close to the Nishimori line because
the transition line has a strong curvature. One needs to change simultaneously
$p$ and $T$, thus we do not have any more correlated samples and
it is much more time consuming to locate the maximum of the central charge
(as is the case on the Nishimori line).
Fig.~\ref{phd} shows the phase diagram obtained from
the maximum of the central charge. Within error bars, our results agree
with the phase diagram Fig.~7 determined in Ref.~\cite{MC} by a different
method. In Fig.~\ref{ccfp} we plot the
corresponding central charge obtained from three-point fits with
$L=3,4,5$ and $L=4,5,6$. Since the difference of the central charge
between the pure Ising model and the Nishimori point is very small,
we employed a parameter $X$ defined as follows:
\beq
X={c_{\rm eff}(p)-c_{\rm eff}(p=0.109) \over c_{\rm eff}(p=0)-c_{\rm eff}(p=0.109)} \; ,
\label{label8.1}
\eeq
similar to the one defined in the study of dilution, see
eq.~(\ref{Xdef}). From this plot, we can see that by increasing the
lattice size, the attractive fixed point is the pure Ising model
(since the variation of the central charge increases, starting from the
Nishimori point). This is consistent with a flow from the
Nishimori point to the pure Ising model and moreover the absence of additional
fixed points along this line.

\subsection{The zero-temperature fixed point} 

\label{secZeroT}

In this subsection, we study the zero-temperature fixed
point. First we present results for the location of the fixed point
at $T=0$. This was already determined with high precision for the $\pm J$
disorder case by Wang et al.\ \cite{Preskill} and Amoruso and Hartmann
\cite{Hartmann}. The result obtained by these groups is
$p^0_c \simeq 0.103$, indicating a re-entrance of the paramagnetic
phase. For comparison, we repeated these measurements, obtaining
compatible results. We further considered the Gaussian case for which the
same re-entrance effect is obtained.

\begin{figure}
\begin{center}
\epsfxsize=400pt{\epsffile{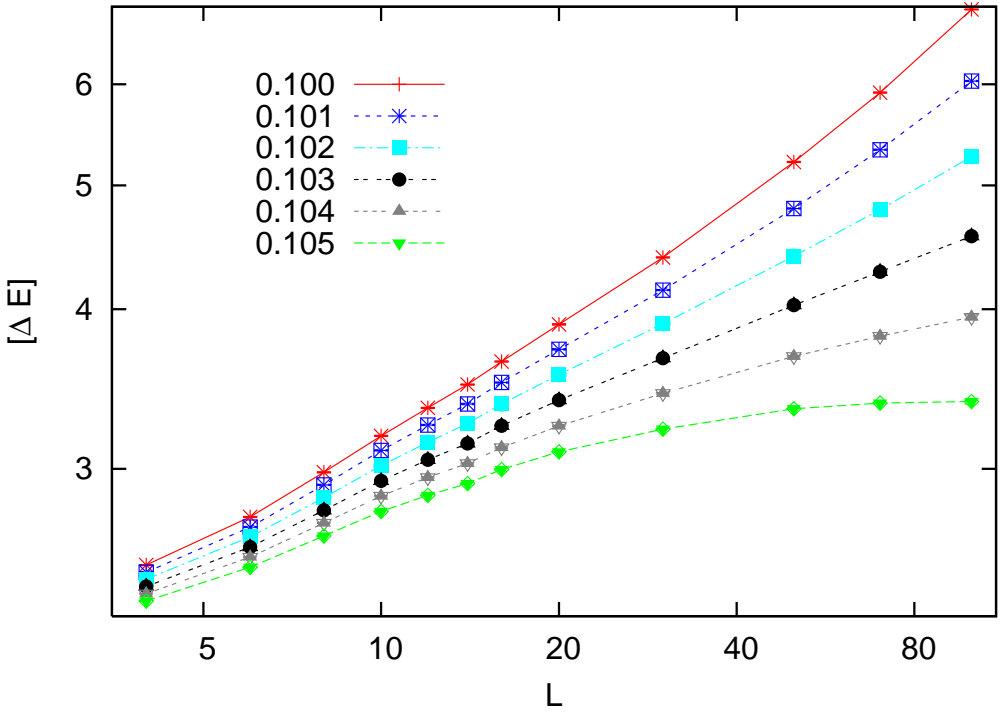}}
\end{center}
\protect\caption[2]{\label{ztdwpm} Domain-wall energy $[\Delta E]$ as a
function of the system size $L$ for the $\pm J$ disorder at $T=0$ and
for $p=0.100, \ldots, 0.105$.}
\end{figure}

\begin{figure}
\begin{center}
\epsfxsize=400pt{\epsffile{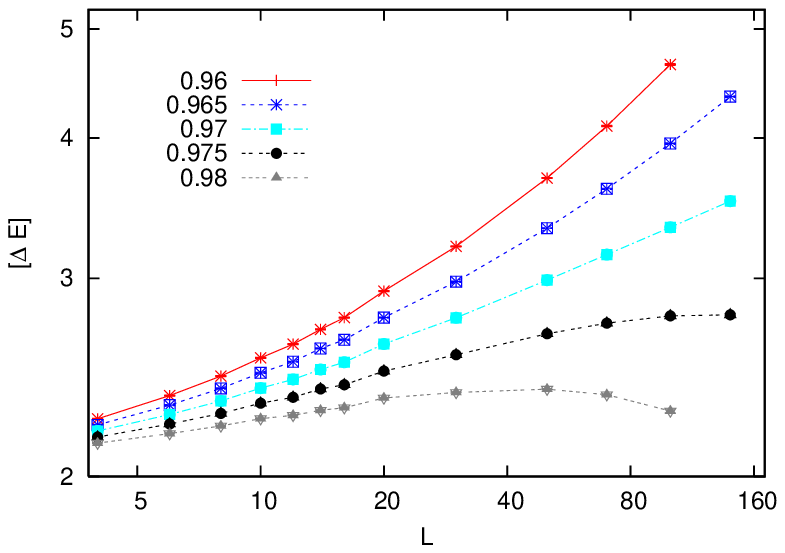}}
\end{center}
\protect\caption[2]{\label{ztdwg} Domain-wall energy $[\Delta E]$ as a
function of the system size $L$ for the Gaussian disorder at $T=0$ and
for $\sigma =0.96, \ldots, 0.98$.}
\end{figure}

The simulations were performed on square lattices with free boundary
conditions along one direction
and periodic/antiperiodic conditions in the other
direction. We measure the difference of energy between the periodic
case and the antiperiodic case
\beq
\Delta E = E_p -E_a \; .
\label{label8.2}
\eeq
This difference of energy corresponds to the energy associated to a
domain wall in the system. 
We expect that after averaging over disorder, the domain-wall energy is
characterized by some size exponent $\rho$
\beq
[ \Delta E ] \propto L^\rho \; .
\label{label8.3}
\eeq
This quantity (as well as the one associated to the width of the
distribution of domain-wall energies) was computed with high precision
by Amoruso and Hartmann \cite{Hartmann} on very large lattices, up to
$L=700$, by using a minimum-weight perfect matching algorithm. In
Fig.~\ref{ztdwpm}, we present our results on smaller sizes (up to $L=100$)
but with much larger statistics, while employing the same type of
algorithm \cite{blossom4}. We simulated 1 million samples for each $p$
and $L < 100$ and 0.5 million samples for $L=100$, compared to
$30~000$ samples in \cite{Hartmann}. The reason for desiring even
better precision
is to be able to observe a crossover in the location of the fixed
point, see the discussion of this point in the following
subsection. In Fig.~\ref{ztdwpm}, we clearly distinguish a scaling in
function of the size $L$ close to $p \simeq 0.103$ for large $L$, in
perfect agreement with the previous results
\cite{Preskill,Hartmann}. We also note that this power-law behaviour
is apparent only for sizes $L > 20$. In Fig.~\ref{ztdwg}, we present
the same quantity for the Gaussian distribution of disorder. Here
again we observe a power law of $[\Delta E]$ with size $L$ close to $\sigma
\simeq 0.97$ to be compared to the result (\ref{sigmaCritVal})
$\sigma_c \simeq 0.97945$ on
the Nishimori line. For the Gaussian disorder, we have data up to size
$L=140$ and the number of samples is again 1 million for each value of
$\sigma$ and $L$ except for $L=140$ where we have 0.5 million samples.

Thus in both cases, we observe a re-entrance of the paramagnetic phase. 

\subsection{Magnetic exponent at zero temperature}

In order to characterize the fixed point at zero temperature
we measure the magnetic exponent in the same manner as on the Nishimori
line. Specifically, we use measurements of the spin-spin correlation
functions which provide a direct estimate of the magnetic
exponents and allow us to study higher moments.

\begin{figure}
\begin{center}
%\leavevmode
\epsfxsize=400pt{\epsffile{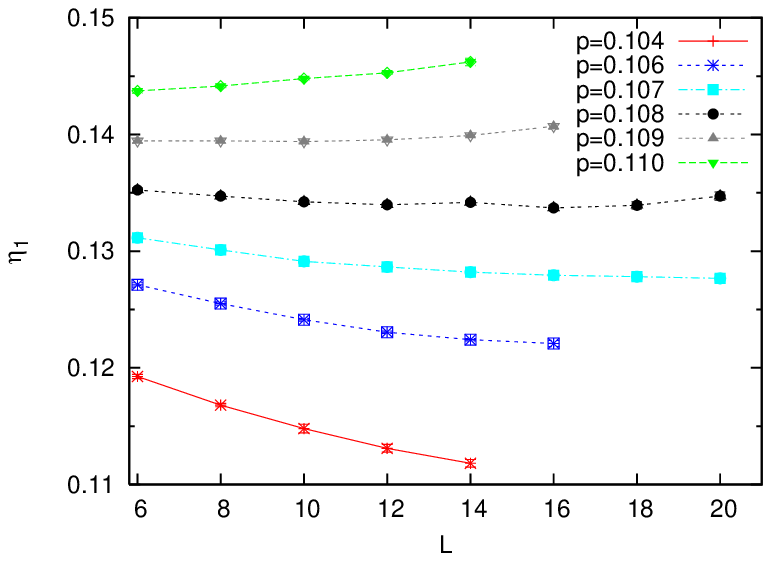}}
\end{center}
\protect\caption[2]{\label{mgaus01} Effective magnetic exponent $\eta_1$
vs.\ $L$ for the $\pm J$ disorder case at $T=0$.}
\end{figure}

\begin{figure}
\begin{center}
%\leavevmode
\epsfxsize=400pt{\epsffile{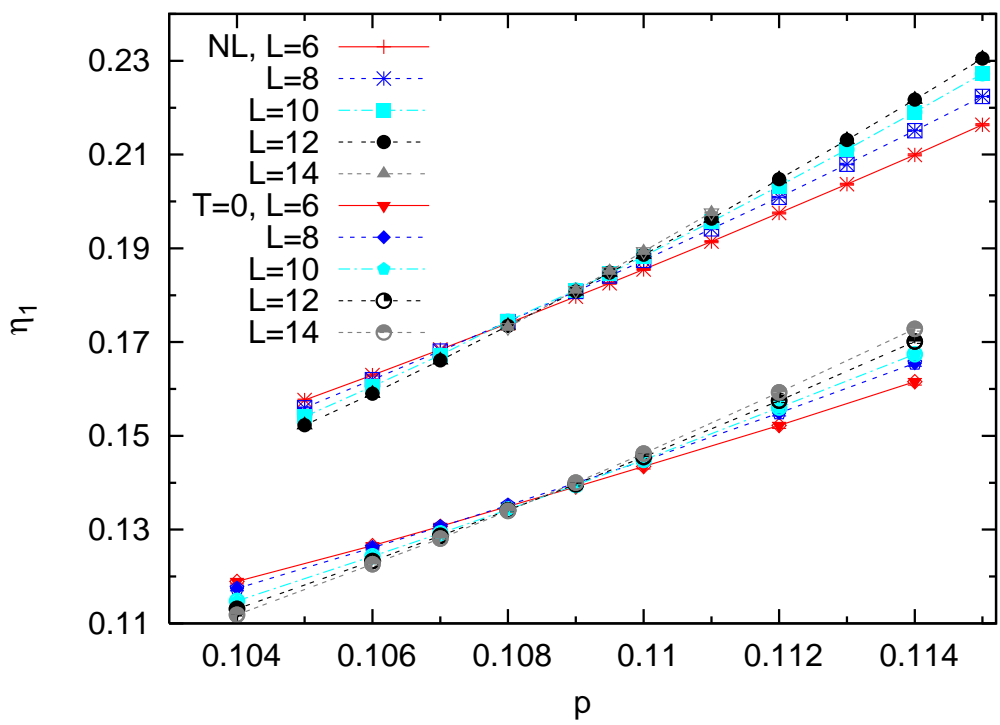}}
\end{center}
\protect\caption[2]{\label{mgaus02} Effective magnetic exponent $\eta_1$
vs.\ $p$ for $\pm J$ disorder case on the Nishimori line and at $T=0$.}
\end{figure}

Fig.\ \ref{mgaus01} shows the values of $\eta_1$, obtained from
the measured spin-spin correlation functions with a fit to the form
eq.\ (\ref{fitcf1}). We see in this figure that $\eta_1$ is constant
close to $p=0.107-0.108$, at least for the largest sizes that we can
reach, $L=20$. For larger sizes, we expect that this value can
still decrease, see the discussion below.
Next, in Fig.\ \ref{mgaus02}, we compare the magnetic exponent (or
more precisely $\eta_1 = 2 x_h$) obtained for the $\pm J$ distribution
of disorder, both on the Nishimori line and at $T=0$. In this figure,
one can see that the critical point at $T=0$, denoted by $p_c^0$ is
very close to $p_c$, the critical point on the Nishimori line. The
best measurements yields a value of $p_c^0$ slightly smaller than $p_c$,
close to $0.108$. This value is far from the one obtained in the
previous section, {\it i.e.} $p_c^0 = 0.103$. We explain this
difference by the existence of strong finite-size corrections, a
situation which is frequent for two-dimensional spin glasses, see
\cite{HM}. These finite-size corrections can also be observed
directly in Fig.\ \ref{mgaus01}. For small sizes, $6\le L \le10$, the
effective exponent $\eta_1$ is constant for $p \simeq 0.109$ and only
for this value of $p$. For $10\le L \le14$, it is almost constant for $p
\simeq 0.108$. By increasing $L$, the estimate for $p_c^0$ will continue
to decrease. For the largest size that we simulated, $p_c^0$ is estimated
close to $0.107$. Presumably, this value will still decrease with increasing
size. In the discussion in the previous subsection, we had already
observed that a power-law behaviour is apparent only for $L > 20$.

\begin{figure}
\begin{center}
\epsfxsize=400pt{\epsffile{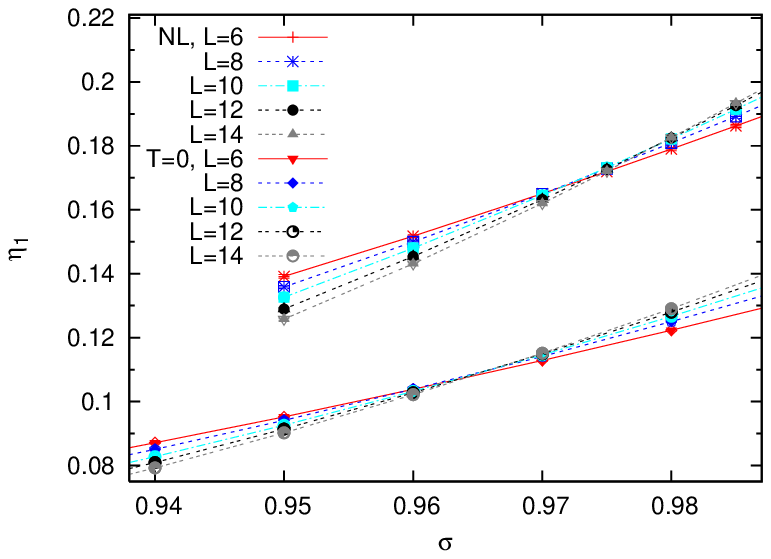}}
\end{center}
\protect\caption[2]{\label{mgaus03} Effective magnetic exponent $\eta_1$
vs.\ $\sigma$ for Gaussian disorder on the Nishimori line and at $T=0$.}
\end{figure}

Fig.\ \ref{mgaus03} shows $\eta_1$ for the Gaussian distribution
of disorder on the Nishimori line and at $T=0$. Here again, we clearly
observe a small re-entrance, with $\sigma_c^0 \simeq
0.97$ compared to $\sigma_c \simeq 0.97945$ (see (\ref{sigmaCritVal}))
on the Nishimori line. 
Contrary to the $\pm J$ disorder case, there are very small finite-size
effects. For $L \simeq 14$, one already obtains the same result for $\sigma_c^0$
as by domain-wall measurements on much bigger systems. 

Finally, let us compare the value of the magnetic
exponents for the two types of disorder at $T=0$. For the Gaussian
disorder, one obtains $\eta_1 = 0.11-0.12$, see Fig.\ \ref{mgaus03}.
The $\pm J$ disorder case is less clear. In Fig.~\ref{mgaus02},
as discussed above, one still has strong finite-size
corrections. If one considers the asymptotic point
$p_c^0 \simeq 0.103$ (compare section \ref{secZeroT}),
then the corresponding $\eta_1$ will be close to $0.11$, thus
rather close to the one of the $\pm J$ disorder case. Still, this  is not
sufficient to conclude that we have universality at $T=0$. Indeed, at
$T=0$, the type of disorder influences the degeneracy of the ground
states. For the Gaussian distribution of disorder, each correlation
function (before averaging over disorder) will be equal to $\pm 1$.
Thus the even moments will always be equal to one, while the odd
moments will be equal to the first moment. Then, one obtains the
general result, valid as long the ground state is unique (up to a
global symmetry), $\eta_{2n+1} = \eta_1$ for any $n$ and $\eta_{2n}=
0$.  On the contrary, for the $\pm J$ disorder case, we have a huge
degeneracy of the ground states. A direct determination of the
exponents $\eta_i$ for $L=20$ and $p=0.107$ yields:
\bea
\eta_1 = 0.1277 \pm  0.0004 \; &;& \; \eta_2 = 0.0593 \pm  0.0002 \nonumber \\
\eta_3 = 0.1369 \pm  0.0004 \; &;& \; \eta_4 = 0.0719 \pm  0.0002 \nonumber \\
\eta_5 = 0.1412 \pm  0.0004 \; &;& \; \eta_6 = 0.0776 \pm  0.0002 \nonumber \\
\eta_7 = 0.1438 \pm  0.0004 \; &;& \; \eta_8 = 0.0810 \pm  0.0002 \, .
\eea
The value $p=0.107$ is chosen even if it is not exactly at the
asymptotic critical point $p_c^0 \simeq 0.103$ since for this value
we have measurements for large sizes. We also computed the same
exponents for smaller $p$ and smaller $L$. The change in the exponents
is very small. We thus conclude that for the $\pm J$ disorder case,
all moments are different.

\section{Summary and conclusions}

In this paper we have performed an extensive study of the random-bond
Ising model with a particular emphasis on non-trivial fixed points.
Concerning the Nishimori point, which is  located at the intersection 
of the Ferro-Para critical line and the Nishimori line, we have used
domain-wall free energy computations to provide an accurate estimate 
for the location of this multicritical point. This has been done for both, the binary
$\pm J$ distribution as well as for the Gaussian distribution.
Our numerical results show that a conjecture for the location of
the Nishimori point based on a duality argument \cite{NN} yields
only a very good approximation, but is not exact.

Next, we estimated $\nu = 1.48(3)$ for $\pm J$ disorder and $\nu =
1.52(3)$ for Gaussian disorder. This agrees with other recent
estimates for $\pm J$ disorder, namely $\nu = 1.50(3)$ on the square
lattice \cite{MC}, and $\nu = 1.49(2)$ on the triangular and honeycomb
lattices \cite{deQTH}. All these results are consistent with a
universal value $\nu \approx 1.50$ for the Nishimori point. We have
obtained further accurate results for the exponents $\eta_1, \ldots,
\eta_8$ for the moments of the spin-spin correlation functions. The
estimates for $\pm J$ disorder (\ref{valExp20}) and Gaussian disorder
(\ref{valExp14}) are not only very close to each other, but also to
recent estimates on the triangular and honeycomb lattices
\cite{deQTH}.  Our analysis of the central charge $c$ at the Nishimori
point for both types of disorder is also consistent with a universal
value $c = 0.464 \pm 0.004$ \cite{HPP}.
All these results\footnote{Essentially the same results are found \cite{LQ06}
if the approximate value $p_c \approx 0.110028$ \cite{NN} is used
instead of the numerically exact value (\ref{pcrit1}).}
suggest a single universality class of the Nishimori point in the
two-dimensional random-bond Ising model, and definitely exclude
percolation as the possible universality class for this point.

We have also considered a probability distribution for the bonds with
dilution, in which some of the coupling constants are zero.
In the purely diluted case (a distribution containing
only positive or zero bonds) there is only one non-trivial fixed point,
namely the zero-temperature percolation point, apart from
the critical point of the pure system \cite{YeSt79}.
We have confirmed that this percolation point is unstable against
the Nishimori point
if one moves within the intersection of the critical transition
surface and the Nishimori manifold by considering both, dilution and
$\pm J$ couplings, and that these two points (percolation and Nishimori)
are the only fixed points within this intersection line. 
On the other hand, going off the Nishimori line but staying within
the critical transition line we confirm that the Nishimori
point is unstable in favour the pure Ising model fixed point. All these
results are obtained by studying the crossover of the effective
central charge, or first Lyapunov exponent, in strips of increasing
width.

Finally, we have analyzed the critical point at zero temperature
corresponding to the ferro-para transition in the model without dilution
for both the binary and Gaussian distributions. Our numerical analysis
confirms the strict re-entrance of the ferromagnetic
phase \cite{Preskill,Hartmann}. We have also investigated the
criticality of the zero-temperature fixed point and argued that it is,
as for the finite-temperature Nishimori point, different from
percolation. The results obtained in this paper raise the question
of the apparent hierarchy of fixed points for strongly disordered systems,
namely in our case the pure Ising transition, percolation, the Nishimori
point and the zero-temperature point, all corresponding to different (and
certainly non-unitary) conformal field theories with values for the
central charge and critical exponents which are extremely close to
each other but nevertheless different. Future analytical efforts will
be needed to understand such effects
of strong disorder in two-dimensional classical statistical systems.

%\noindent{\large\bf Acknowledgments} 

%\vspace*{0.7 true cm}
%
%\newpage
%\small


\begin{thebibliography}{99}


\bibitem{DDSL}
Vik. S. Dotsenko and Vl. S. Dotsenko,
\newblock {\it Sov.\ Phys.\ JETP Lett.}\ {\bf 33} (1981) 37; 
\newblock {\it Adv.\ Phys.}\ {\bf 32} (1983) 129;
B. N. Shalaev,
\newblock {\it Sov. Phys. Solid State}~{\bf 26} (1984) 1811;
A. W. W. Ludwig,
\newblock {\it Nucl. Phys. }~B~{\bf 330} (1990) 639

\bibitem{N}
H. Nishimori,
\newblock {\it J.\ Phys.\ C: Solid State Phys.}\ {\bf 13} (1980) 4071;
\newblock {\it Prog.\ Theor.\ Phys.}~{\bf 66} (1981) 1169

\bibitem{ON}
Y. Ozeki and H. Nishimori,
\newblock {\it J. Phys. A: Math. Gen.}~{\bf 26} (1993) 3399

\bibitem{LG}
A. Georges and P. Le Doussal,
\newblock {\it unpublished Preprint} (1988); 
P.~Le Doussal and A.~B.\ Harris, {\it Phys.\ Rev.\ Lett.}~{\bf 61} (1988) 625;
{\it Phys. Rev.}~B~{\bf 40} (1989) 9249

\bibitem{CF}
S. Cho and M. P. A. Fisher,
\newblock {\it Phys. Rev.}~B {\bf 55} (1997) 1025

\bibitem{GRL}
I. A. Gruzberg, N. Read and A.~W.~W.~Ludwig,
\newblock {\it Phys. Rev.}~B {\bf 63} (2001) 104422

\bibitem{CRKHAL} J.\ T.\ Chalker, N.\ Read, V.\ Kagalovsky, B.\ Horovitz, Y.\ Avishai and A.\ W.\ Ludwig
\newblock {\it Phys. Rev.}~B {\bf 65} (2001) 012506

\bibitem{MENMD06} A.\ Mildenberger, F.\ Evers, R.\ Narayanan, A.~D.\ Mirlin
              and K.\ Damle, {\it Phys.\ Rev.}\ B {\bf 73} (2006) 121301(R)
	      
\bibitem{Sourlas} N.\ Sourlas, {\it Europhys.\ Lett.}\ {\bf 25} (1994) 159;
               {\it Preprint} cond-mat/9811406

\bibitem{NishCod} H.\ Nishimori, {\it Physica} A {\bf 205} (1994) 1;
                  {\it Physica} A {\bf 315} (2002) 243

\bibitem{Iba} Y.\ Iba, {\it J.\ Phys.\ A: Math.\ Gen.}\ {\bf 32} (1999) 3875

\bibitem{Preskill}
C. Wang, J. Harrington and J. Preskill,
\newblock {\it Annals Phys.}~{\bf 303} (2003) 31

\bibitem{Z}
M. R. Zirnbauer,
\newblock {\it J. Math. Phys.}~{\bf 37} (1996) 4986; 
A. Altland and M. R. Zirnbauer,
\newblock {\it Phys. Rev.}~B {\bf 55} (1997) 1142

\bibitem{GRLC}
I. A. Gruzberg, N. Read and A.~W.~W.~Ludwig,
\newblock {\it Phys. Rev. Lett.}~{\bf 82} (1999) 4524;
J. Cardy,
\newblock {\it Phys. Rev. Lett.}~{\bf 84} (2000) 3507


\bibitem{HPP} A.\ Honecker, M.\ Picco and P.\ Pujol,
\newblock {\it Phys. Rev. Lett.}~{\bf 87} (2001) 047201

\bibitem{MM}
W. L. McMillan,
\newblock {\it Phys. Rev.}~B {\bf 29} (1984) 4026

\bibitem{SA}
R. R. P. Singh and J. Adler,
\newblock {\it Phys. Rev.}~B {\bf 54} (1996) 364

\bibitem{AQdS}
F. D. A. Ar\~ao Reis, S. L. A. de Queiroz and R. R. dos Santos,
\newblock {\it Phys. Rev.}~B {\bf 60} (1999) 6740

\bibitem{MC}
F. Merz and J.~T. Chalker,
\newblock {\it Phys. Rev.}~B {\bf 65} (2002) 054425

\bibitem{MC2} F.\ Merz and J.~T.\ Chalker, 
%{\it Negative Scaling Dimensions and
%              Conformal Invariance at the Nishimori Point in the $\pm J$
%              Random-Bond Ising Model}, % {\it Preprint} cond-mat/0201137
\newblock {\it Phys.\ Rev.}~B~{\bf 66} (2002) 054413

\bibitem{NN}
H. Nishimori and K. Nemoto,
\newblock  {\it J. Phys. Soc. Jpn.}~{\bf 71} (2002) 1198;
H.\ Nishimori, {\it Preprint} cond-mat/0602453

\bibitem{Grinstein}
G. Grinstein, C. Jayaprakash and M. Wortis,
\newblock {\it Phys. Rev.}~B {\bf 19} (1979) 260

\bibitem{Freund}
H. Freund and P. Grassberger,
\newblock {\it J. Phys. A: Math. Gen.}~{\bf 22} (1989) 4045

\bibitem{Bendish}
J. Bendish, U. Derigs and A. Metz,
\newblock {\it Discrete Applied Mathematics}~ {\bf 52} (1994) 139

\bibitem{KaRi} N.\ Kawashima and H.\ Rieger, 
%{\it Finite Size Scaling Analysis of
% Exact Ground States for $\pm J$ Spin Glass Models in Two
%             Dimensions}, 
\newblock {\it Europhys.\ Lett.}\ {\bf 39} (1997) 85

\bibitem{MB} G.\ Migliorini and A.\ N.\ Berker, 
\newblock {\it Phys.\ Rev.}~B~{\bf 58} (1998) 426

\bibitem{BGP} J.~A.\ Blackman, J.~R.\ Gon\c{c}alves and J.\ Poulter, 
%{\it Properties
%              of the Two-Dimensional Random-Bond $\pm J$ Ising Spin Glass},
\newblock {\it Phys.\ Rev.}~E~{\bf 58} (1998) 1502

\bibitem{Hartmann}
C. Amoruso and A. K. Hartmann,
\newblock {\it Phys. Rev.}~B {\bf 70} (2004) 134425

\bibitem{NS}
C.~M. Newman and D.~L. Stein,
\newblock {\it Phys. Rev. Lett.}~{\bf 84} (2000) 3966;
\newblock {\it Commun.\ Math.\ Phys.}\ {\bf 224} (2001) 205 % cond-mat/0103395.

\bibitem{OzekiNishimori} 
Y.\ Ozeki and H.\ Nishimori, 
%{\it Phase Diagram of the
%$\pm J$ Ising Model in Two Dimensions}, 
\newblock {\it J.\ Phys.\ Soc.\ Jpn.}\ {\bf 56} (1987) 3265

\bibitem{OzekiIto} Y.\ Ozeki and N.\ Ito, 
%{\it Multicritical Dynamics for the
%              $\pm J$ Ising Model}, 
\newblock {\it J.\ Phys.\ A: Math.\ Gen.}\ {\bf 31} (1998) 5451

\bibitem{MNN03} J.~M.\ Maillard, K.\ Nemoto and H.\ Nishimori, 
%{\it Symmetry,
%              Complexity and Multicritical Point of the Two-Dimensional Spin
%              Glass}, 
\newblock {\it J.\ Phys.\ A: Math.\ Gen.}\ {\bf 36} (2003) 9799

\bibitem{MS95} G.\ Mussardo and P.\ Simonetti, {\it Phys.\ Lett.}\ B {\bf 351}
              (1995) 515

\bibitem{CHMP} D.~C.\ Cabra, A.\ Honecker, G.\ Mussardo and P.\ Pujol,
          {\it J.\ Phys.\ A: Math.\ Gen.}\ {\bf 30} (1997) 8415

\bibitem{Night} M.~P.\ Nightingale, pp.\ 287-351 in: V.\ Privman (ed.),
              {\it Finite Size Scaling and Numerical Simulations of
              Statistical Physics}, World Scientific, Singapore (1990)

\bibitem{YeSt79} J.~M.\ Yeomans and R.~B.\ Stinchcombe,
              {\it J.\ Phys.\ C: Solid State Phys}.\ {\bf 12} (1979) 347

\bibitem{StAh} D.\ Stauffer and A.\ Aharony, {\it Introduction to Percolation
              Theory}, 2nd edition, Taylor \& Francis, London (1994)
              

\bibitem{Sor} E.~S.\ S{\o}rensen,
% {\it Logarithmic Corrections to the RG Flow
%              for the Two-Dimensional Bond Disordered Ising Model},
              {\it Preprint} cond-mat/0006233


\bibitem{HJPP} A.\ Honecker, J.~L.\ Jacobsen, M.\ Picco and P.\ Pujol, 
%{\it
%              Nishimori Point in Random-Bond Ising and Potts Models in 2D},
%              Proceedings of the NATO Advanced Research Workshop on
\newblock {\it Statistical Field Theories}, Como, 18-23 June 2001,
             eds.\ A.\ Cappelli, G.\ Mussardo, Kluwer Academic Publishers,
             Dordrecht (2002) 251-261 [{\it Preprint} cond-mat/0112069].

\bibitem{central} H. W. J. Bl\"ote, J. L. Cardy and M. P. Nightingale,
\newblock {\it Phys.\ Rev.\ Lett.}\ {\bf 56} (1986) 742;
I.\ Affleck,
\newblock {\it Phys.\ Rev.\ Lett.}\ {\bf 56} (1986) 746

\bibitem{JC}
J.~L. Jacobsen and J.~L. Cardy, 
{\it Nucl. Phys.}~B~{\bf 515} (1998) 701

\bibitem{Cardy} J.\ Cardy, {\it Scaling and Renormalization in Statistical
              Physics}, Cambridge Lecture Notes in Physics 5, Cambridge
              University Press (1996)

\bibitem{WK74} K.~G.\ Wilson and J.\ Kogut,
              {\it Phys.\ Rep.}\ {\bf 12} (1974) 75

\bibitem{QSnew} S.~L.~A.\ de Queiroz and R.~B.\ Stinchcombe, 
%{\it
%              Correlation-Function Distributions at the Nishimori Point of
%              Two-Dimensional Ising Spin Glasses}, 
\newblock {\it Phys.\ Rev.}~B~{\bf 68} (2003) 144414

\bibitem{cardyr} J.~L.~Cardy, in {\it Phase Transitions and Critical Phenomena},
                 Vol. 11,  edited by C. Domb and J. Lebowitz,
		 Academic Press, London (1987)

\bibitem{Binder81} K.\ Binder, {\it Z.\ Phys.}\ B {\bf 43} (1981) 119

\bibitem{LB00} D.~P.\ Landau and K.\ Binder, {\it A Guide to Monte Carlo
     Simulations in Statistical Physics}, Cambridge University Press (2000)

\bibitem{NK}
H. Nishimori,
\newblock {\it J. Phys. Soc. Jpn.}~{\bf 55} (1986) 3305;
H. Kitatani,
\newblock {\it J. Phys. Soc. Jpn.}~{\bf 61} (1992) 4049

\bibitem{JP} J.~L.\ Jacobsen and M.\ Picco, 
\newblock {\it Phys.\ Rev.}~E~{\bf 65} (2002) 026113

\bibitem{Ludwig} A.~W.~W.\ Ludwig, 
\newblock {\it Nucl.\ Phys.}~B~{\bf 285} (1987) 97

\bibitem{DPP} V.~Dotsenko, M.\ Picco and P.\ Pujol,  
\newblock {\it Nucl.\ Phys.}\ B {\bf 455} (1995) 701

\bibitem{blossom4} W.~Cook and A.~Rohe, {\it INFORMS Journal on
Computing}~{\bf 11} (1999) 138

\bibitem{HM} A.~K.\ Hartmann and M.~A.\ Moore, 
\newblock {\it Phys.\ Rev.\ Lett.}\ {\bf 90} (2003) 127201

\bibitem{deQTH} S.~L.~A.\ de Queiroz, {\it Phys.\ Rev.}~B~{\bf 73} (2006) 064410

\bibitem{LQ06} J.~C.\ Lessa and S.~L.~A.\ de Queiroz,
         {\it Preprint} cond-mat/0605659

\end{thebibliography}
\end{document}